\documentclass[apj]{emulateapj}
\usepackage{graphicx}
\usepackage{rotating}
\usepackage{longtable}

\newcommand{\myemail}{andrea.pastorello@oapd.inaf.it}
\newcommand{\rsun}{R$_\odot$}

\shorttitle{SN 2009ip, is this the end?}
\shortauthors{Pastorello et al.}

\begin{document}

\title{Interacting Supernovae and Supernova Impostors.  \\  SN 2009ip, is this the end?}

\author{A. Pastorello\altaffilmark{1}\altaffilmark{$\star$}, E. Cappellaro\altaffilmark{1}, C. Inserra\altaffilmark{2},   S. J. Smartt\altaffilmark{2},
G. Pignata\altaffilmark{3}, S. Benetti\altaffilmark{1}, S. Valenti\altaffilmark{4,5}, M. Fraser\altaffilmark{2}, K. Tak\'ats\altaffilmark{3,6},
S. Benitez\altaffilmark{7}, M. T. Botticella\altaffilmark{8}, J. Brimacombe\altaffilmark{9}, F. Bufano\altaffilmark{3}, F. Cellier-Holzem\altaffilmark{10},
M. T. Costado\altaffilmark{11}, G. Cupani\altaffilmark{12}, I. Curtis\altaffilmark{13},
N. Elias-Rosa\altaffilmark{14}, M. Ergon\altaffilmark{15}, J. P. U. Fynbo\altaffilmark{16}, F.-J. Hambsch\altaffilmark{17}, 
M. Hamuy\altaffilmark{18}, A. Harutyunyan\altaffilmark{19}, K. M. Ivarson\altaffilmark{20}, E. Kankare\altaffilmark{21}, J. C. Martin\altaffilmark{22},
R. Kotak\altaffilmark{2}, A. P. LaCluyze\altaffilmark{20}, K. Maguire\altaffilmark{23}, S. Mattila\altaffilmark{24}, J. Maza\altaffilmark{18},
M. McCrum\altaffilmark{2}, M. Miluzio\altaffilmark{25}, H. U. Norgaard-Nielsen\altaffilmark{16}, M. C. Nysewander\altaffilmark{20},
P. Ochner\altaffilmark{1}, Y.-C. Pan\altaffilmark{23},  M. L. Pumo\altaffilmark{1}, D. E. Reichart\altaffilmark{20}, T. G. Tan\altaffilmark{26},
S. Taubenberger\altaffilmark{7}, L. Tomasella\altaffilmark{1}, M. Turatto\altaffilmark{1}, and D. Wright\altaffilmark{2}}

\altaffiltext{1}{INAF-Osservatorio Astronomico di Padova, Vicolo dell'Osservatorio 5,  35122 Padova, Italy}
\altaffiltext{$\star$}{\myemail}
\altaffiltext{2}{Astrophysics Research Centre, School of Mathematics and Physics, Queen's University Belfast, Belfast BT7 1NN, United Kingdom}
\altaffiltext{3}{Departamento de Ciencias Fisicas, Universidad Andres Bello, Avda. Republica 252, Santiago, Chile}
\altaffiltext{4}{Las Cumbres Observatory Global Telescope Network, Inc. Santa Barbara, CA 93117, USA}
\altaffiltext{5}{Department of Physics, University of California Santa Barbara, Santa Barbara, CA 93106-9530, USA}
\altaffiltext{6}{Department of Optics $\&$ Quantum Electronics, University of Szeged, D\'om t\'er 9, Szeged, H-6720, Hungary}
\altaffiltext{7}{Max-Planck-Institut f\"ur Astrophysik, Karl-Schwarzschild-Str. 1, 85741 Garching, Germany}
\altaffiltext{8}{INAF-Osservatorio Astronomico di Capodimonte, Salita Moiariello 16, I-80131 Napoli, Italy}
\altaffiltext{9}{Coral Towers Observatory, Coral Towers, Esplanade, Cairns 4870, Australia}
\altaffiltext{10}{Laboratoire de Physique Nucl\'eaire et des Hautes Énergie, Universit\'e Pierre et Marie Curie Paris 6, Universit\'e Paris Diderot, Paris 7, CNRS-IN2P3, 4 place Jussieu, F-75252 Paris Cedex 05, France}
\altaffiltext{11}{Instituto de Astrof\'isica de Andaluc\'ia, CSIC, Apdo 3004, 18080, Granada, Spain}
\altaffiltext{12}{INAF - Osservatorio Astronomico di Trieste, via Tiepolo 11, I-34143 Trieste, Italy}
\altaffiltext{13}{2 Yandra Street, Vale Park, Adelaide, South Australia 5081, Australia}
\altaffiltext{14}{Institut de Ciencies de l'Espai (IEEC-CSIC), Campus UAB, 08193 Bellaterra, Spain}
\altaffiltext{15}{The Oskar Klein Centre, Department of Astronomy, AlbaNova, Stockholm University, 106 91 Stockholm, Sweden}
\altaffiltext{16}{Dark Cosmology Centre, Niels Bohr Institute, Copenhagen University, Juliane Maries Vej 30, 2100 Copenhagen O, Denmark}
\altaffiltext{17}{Vereniging Voor Sterrenkunde, Oude Bleken 12, 2400 Mol, Belgium}
\altaffiltext{18}{Departamento de Astronom\'ia, Universidad de Chile, Casilla 36-D, Santiago, Chile}
\altaffiltext{19}{Telescopio Nazionale Galileo, Fundaci\'on Galileo Galilei - INAF, Rambla Jos\'e Ana Fern\'andez P\'erez, 7, 38712 Bre\~na Baja, TF, Spain}
\altaffiltext{20}{University of North Carolina at Chapel Hill, Campus Box 3255, Chapel Hill, NC 27599-3255, USA}
\altaffiltext{21}{Tuorla Observatory, Department of Physics and Astronomy, University of Turku, Piikki\"o, 21500, Finland}
\altaffiltext{22}{Astronomy/Physics MS HSB 314, One University Plaza Springfield, IL 62730, USA}
\altaffiltext{23}{Department of Physics (Astrophysics), University of Oxford, DWB, Keble Road, Oxford OX1 3RH, Uited Kingdom}
\altaffiltext{24}{Finnish Centre for Astronomy with ESO (FINCA), University of Turku, V\"ais\"al\"antie 20, FI-21500, Piikki\"o, Finland}
\altaffiltext{25}{Department of Astronomy, Padova University, Vicolo dell’Osservatorio 3, I-35122, Padova, Italy}
\altaffiltext{26}{115 Adelma Rd, Dalkeith, Western Australia 6009, Australia}


\begin{abstract}

  We report the results of a 3 year-long dedicated monitoring campaign
  of a restless Luminous Blue Variable (LBV) in NGC 7259. The object,
  named SN 2009ip, was observed photometrically and spectroscopically
  in the optical and near-infrared domains. We monitored a number of
  erupting episodes in the past few years, and increased the density of
  our observations during eruptive episodes. In this paper we present
  the full historical data set from 2009-2012 with multi-wavelength
  dense coverage of the two high luminosity events between August -
  September 2012. We construct bolometric light curves and measure the 
total luminosities of these eruptive or explosive events.  
We label them the 2012a event (lasting $\sim$\,50 days) with a peak of 
$3\times10^{41}$\,ergs$^{-1}$, and the 2012b event (14 day rise time, still
ongoing) with a peak of $8\times10^{42}$ \,ergs$^{-1}$. The latter
event reached  an absolute R-band magnitude of about -18,
comparable to  that of a core-collapse supernova (SN). 
Our historical monitoring has detected  high-velocity spectral features 
($\sim$13000\,km s$^{-1}$) in September 2011, one year before the
current SN-like event. This implies that the detection of such high
velocity outflows cannot, conclusively, point to a core-collapse SN
origin.  We suggest that the initial peak in the 2012a event was
unlikely to be due to a faint core-collapse SN. We propose that the  high intrinsic
luminosity of the latest peak, the variability
  history of SN 2009ip, and the detection of broad spectral lines 
  indicative of high-velocity ejecta are consistent with a pulsational
  pair-instability event, and that the star may have  survived the
  last outburst. The question of the survival of the LBV progenitor
  star  and its future fate remain open issues, only to be
  answered with future monitoring of this historically unique
  explosion. 

\end{abstract}


\section{Introduction}\label{intro}

Luminous Blue Variables (LBVs) are among the most luminous and massive
stars found in late-type galaxies. In a few cases, these stars have
been observed to produce major eruptions that mimic a genuine
supernova (SN) explosion.  For this reason, they gained the label of
SN impostors \citep{van00}.  The discrimination between SN impostors
(i.e. LBV-type eruptions) and type IIn SNe can be ambiguous
\citep[see e.g. the SN 2011ht-like
objects,][]{rom12,mau12a,hum12,kan12,des09,chu04}.

LBVs are observed in the Milky Way, Local Group galaxies
and beyond \citep[e.g.][]{hum94,hum99,mau06,smi11}. They have high
mass-loss rates and frequently show what is known as S-Doradus
variability during which mass-loss is enhanced, possibly due to
temperature changes and ionization balance of atomic species that
drive the wind \citep{2002A&A...393..543V}. Giant eruptions have been
observed during which several solar masses of material can be ejected \citep[e.g.][]{dav97},
and the intrinsic stellar luminosity increases substantially. The
physical mechanism that triggers these giant eruptions is still
unknown. Based on analysis of SN data, 
a link between {\it some} LBVs and SNe IIn has been proposed
\citep[see e.g.][]{kot06,smi06,smi07,tru08,tru09}. 
There is at least one case (SN 2005gl) in 
which a likely LBV has been observed to explode as luminous SNe IIn
\citep{gal07,gal09},  and one other case (SN 2010jl) for which there is a
plausible argument for a massive  progenitor star of a
type IIn SN \citep[M $>$ 30 M$_{\odot}$,][]{2011ApJ...732...63S}. \footnote{We note that eruptive Wolf-Rayet
  stars, producing impostors with a luminosity similar to that of an
  LBV outburst, have later on been observed to explode as He-rich Ibn
  SNe \citep{pasto07,pasto08a,fol07} or hybrid IIn/Ibn events \citep{pasto08b,smi12}.}
 In this context, a remarkable object is SN 1961V in NGC 1058, whose nature
(SN IIn vs. LBV-type eruption) is still controversial \citep[see][and references therein]{chu04b,koc11,van12}.

In an exciting turn of events, a well observed LBV in the spiral
galaxy NGC 7259  (designated as SN 2009ip during a giant outburst in
2009)  has recently been proposed to have finally exploded as a
core-collapse SN \citep[][and references therein]{mau12}. 
The object was first discovered on August 26, 2009 by the CHASE SN Search \citep{maza09} as a faint transient at
$\approx$ 17.9 mag, and was later classified as a SN impostor by a number of teams
\citep{mil09,li09,ber09}. The nature of SN 2009ip was widely discussed in \citet{smi10,smi11} and \citet{fol11}. Through the analysis 
of pre-outburst archival HST images these studies provided robust evidence that the progenitor was a very massive star 
that experienced repeated eruptions typical of the LBV phase.
The main sequence mass of the star estimated by \citet{smi10} was in the range 50-80 M$_\odot$, whilst \citet{fol11} found it to be M$_{ZAMS} >$ 60 M$_\odot$.

\begin{figure}
\includegraphics[angle=0,scale=2.3]{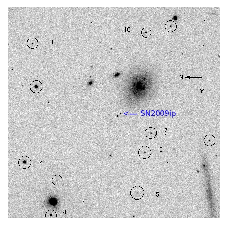}
\caption{SN 2009ip in NGC 7259, and reference stars in the host galaxy field. \label{fig1} }
\end{figure}

Subsequent re-brightenings were announced by the Catalina Real-Time Survey team on October 1, 2010 \citep{dra10}  and, very recently, on July 24, 2012 \citep{dra12},
which were first labeled as new LBV-type outbursts \citep[e.g.][]{fol12}.
However, from the detection of high-velocity
spectral features on September 15 and 16, 2012 \citet{smi12a} first mentioned the possibility that SN 2009ip exploded as a real core-collapse SN
\footnote{We note that after the Smith $\&$ Mauerhan communication  there has been a proliferation of electronic telegrams on this  transient, with different 
interpretations on its nature - SN vs. SN impostor - \citep{marg12a,mart12a,bri12,marg12b,smi12b,leo12,bur12,vin12,pri12,mart12b,gal12,bol12,vin12b,jha12}, although it 
seems that that most authors now favor the SN explosion scenario.}. 
High-cadence optical imaging in the R and I bands monitoring the strong September 2012 re-brightening has
 been presented by \citet{pri12b}. We also note that although no \citep{marg12a,marg12b,cha12,han12} or marginal \citep{cam12,marso12} X-ray and radio 
detections of SN 2009ip were initially reported, a X-ray brightening has been later announced \citep{cam12b}. 

In this paper we present observations of the LBV known as SN 2009ip in NGC
7259 over a period of 3\,yrs including: {\bf i)} data showing the history of erratic variability
starting from August 2009, when the object closely resembled NGC
3432-LBV1 \citep[aka SN 2000ch,][]{wag04,pasto10}, a SN impostor that
experienced multiple energetic outbursts. Our data of SN 2009ip
also include observations of repeated outbursts during the period
May to October 2011 which have not been reported to date; {\bf ii)} the recent
evolution of the LBV as a putative SN.

\section{Observations}

\begin{figure*}
\includegraphics[angle=270,scale=0.65]{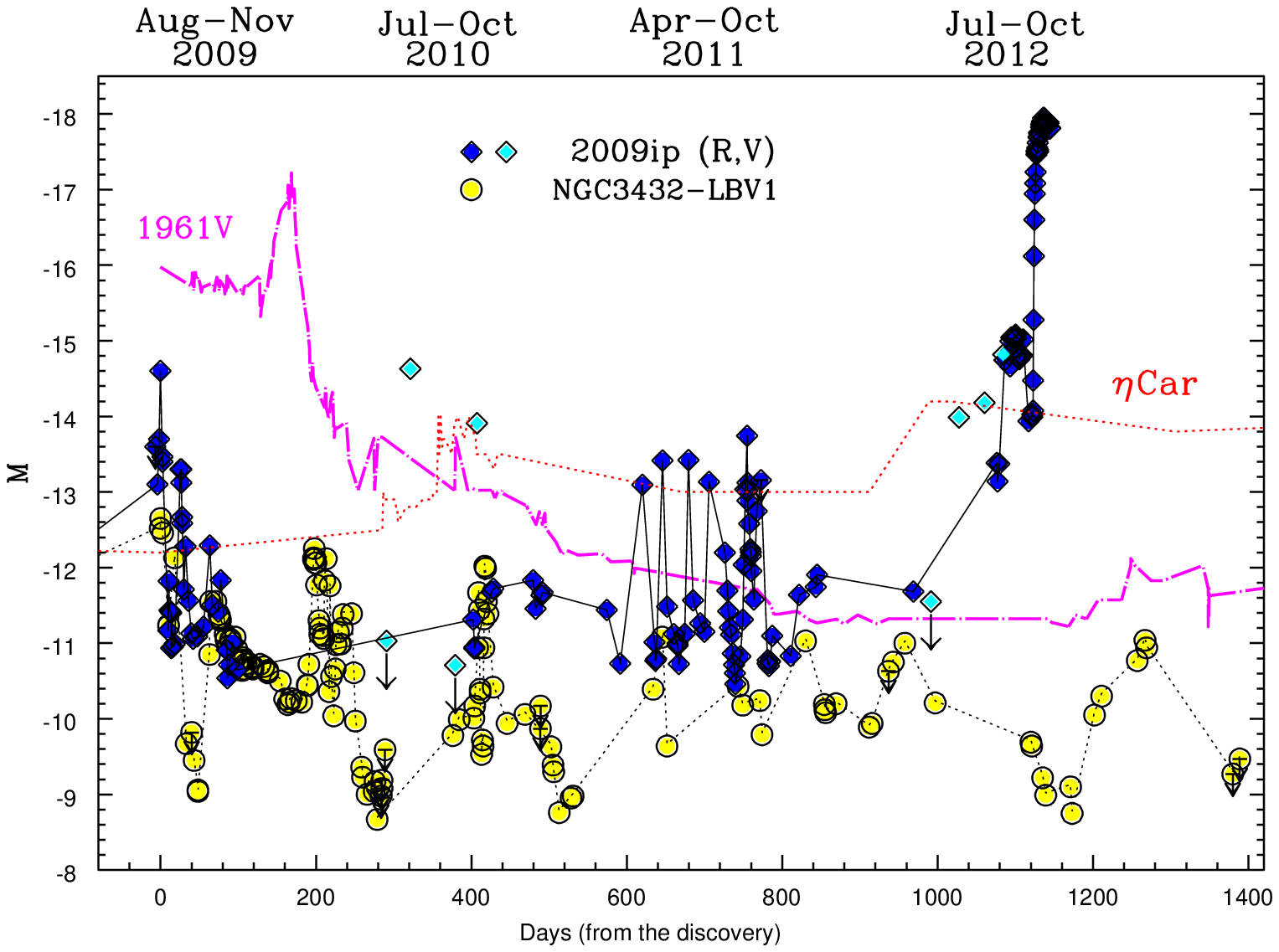}
\caption{R-band absolute light curve of SN 2009ip (blue diamonds) compared with those of the impostor NGC 3432-LBV1 (yellow circles), the debated SN/impostor 1961V
(photographic plate magnitudes, magenta dot-dashed line) and the historical visual light curve of
$\eta$ Carinae during the period 1842-1845 \citep[revised by][red dotted line]{smi11b}. The cyan diamonds represent CRTS V-band measurements \protect\citep[see also][]{dra10,dra12}.
The data showing NGC 3432-LBV1 during the period 2008-2012 are from \citet{pasto10}, plus additional recent unpublished observations  (see Appendix, Table \ref{tab_00ch}).
The epoch 0 of the $\eta$ Carinae light curve is year 1842.213 (UT). The erratic photometric variability 
is a common property of major eruptions of LBVs. \label{fig2} }
\end{figure*}

Three years ago, after the first announcement of the discovery of a
transient in NGC 7259 \citep{maza09}, we initiated an extensive
spectroscopic and photometric monitoring campaign in the optical bands
using a number of telescopes available to our collaboration.  After
about 100 days, the follow-up strategy was relaxed and the photometric
monitoring was limited to the R band. Due to its unpredictable
behavior, we kept up a monitoring campaign of this object during the
following 3 years.

After the recent re-brightening of SN 2009ip announced by the Catalina
Real-Time Survey team on July 24, 2012 \citep{dra12},  we intensified
our  observing cadence and secured multi-color photometry and spectroscopy from the optical to the
near-IR domains. 

In addition, SWIFT optical and ultra-violet observations
have been triggered (PIs: R. Margutti and P. W. A. Roming) and included in our analysis,
particularly to derive a pseudo-bolometric light curve of the 2012 eruptions. 

\subsection{Photometry} \label{photo}

Photometric observations were carried on using a long list of facilities, namely: the 8.2-m Very Large Telescope (VLT) of the European Southern Observatory (ESO) equipped with FORS2 (Cerro Paranal, Chile),
the 3.58-m ESO New Technology Telescope (NTT) equipped with EFOSC2 and SOFI (La Silla, Chile), the 3.58-m Telescopio Nazionale Galileo (TNG) + LRS, the 2-m Liverpool Telescope with RATCam and 
the 2.56-m Nordic Optical Telescope (NOT) + ALFOSC and NOTCam (La Palma, Canary Islands, Spain); the 2-m Faulkes Telescope South + EM02 (Siding Spring Observatory, Australia);
the 0.41-m Panchromatic Robotic Optical Monitoring and Polarimetry Telescopes (Cerro Tololo, Chile); and a group of 0.3 to 0.5-m telescopes in Australia and New Nexico, USA 
(see Table \ref{tab_opt} for details). As mentioned above, the SWIFT satellite plus UVOT secured additional optical and ultra-violet photometry.

The pre-reduction of the optical photometry images was performed using standard IRAF \footnote{IRAF is distributed by the National Optical Astronomy Observatory, 
which is operated by the Association of Universities for Research in Astronomy (AURA) under cooperative agreement with the National Science Foundation.} 
tasks, and these included bias, overscan and flat-field corrections.

The pre-reduction of the NIR photometry images required a few additional steps, since we had to remove from the science images the contribution of the bright NIR sky.
Clear sky images were therefore obtained by median-combining a number of dithered science frames and were then subtracted from the target images. Thereafter, the sky-subtracted science images
were spatially registered and combined in order to improve the signal to noise.

SN 2009ip is located close to a red (R = 18.05 $\pm$ 0.04, R-I = 0.72 $\pm$ 0.05) foreground star, in a remote position North-East of the host galaxy (Figure \ref{fig1}).
Our optical and NIR photometric measurements were performed using the PSF-fitting technique, with the simultaneous fit of the transient and the nearby star. 
A number of reference stars in the SN field were calibrated using observations of  standard fields from the catalog of \citet{lan92},
and used to improve the photometric calibration of the optical photometry of
SN 2009ip in non-photometric nights. The NIR photometry of the reference stars was calibrated agaist the 2-MASS catalogue magnitudes \citep{skr06}.

SWIFT/UVOT data (in the uvw2, uvm2, uvw1, u, b, v bands) were reduced using the heasarc\footnote{NASA’s High Energy Astrophysics Science Archive Research Center.} software. Images obtained on the same epoch were
 co-added, and finally reduced using the prescriptions of \citet{poo08}. 

\begin{figure*}
\epsscale{1}
\plotone{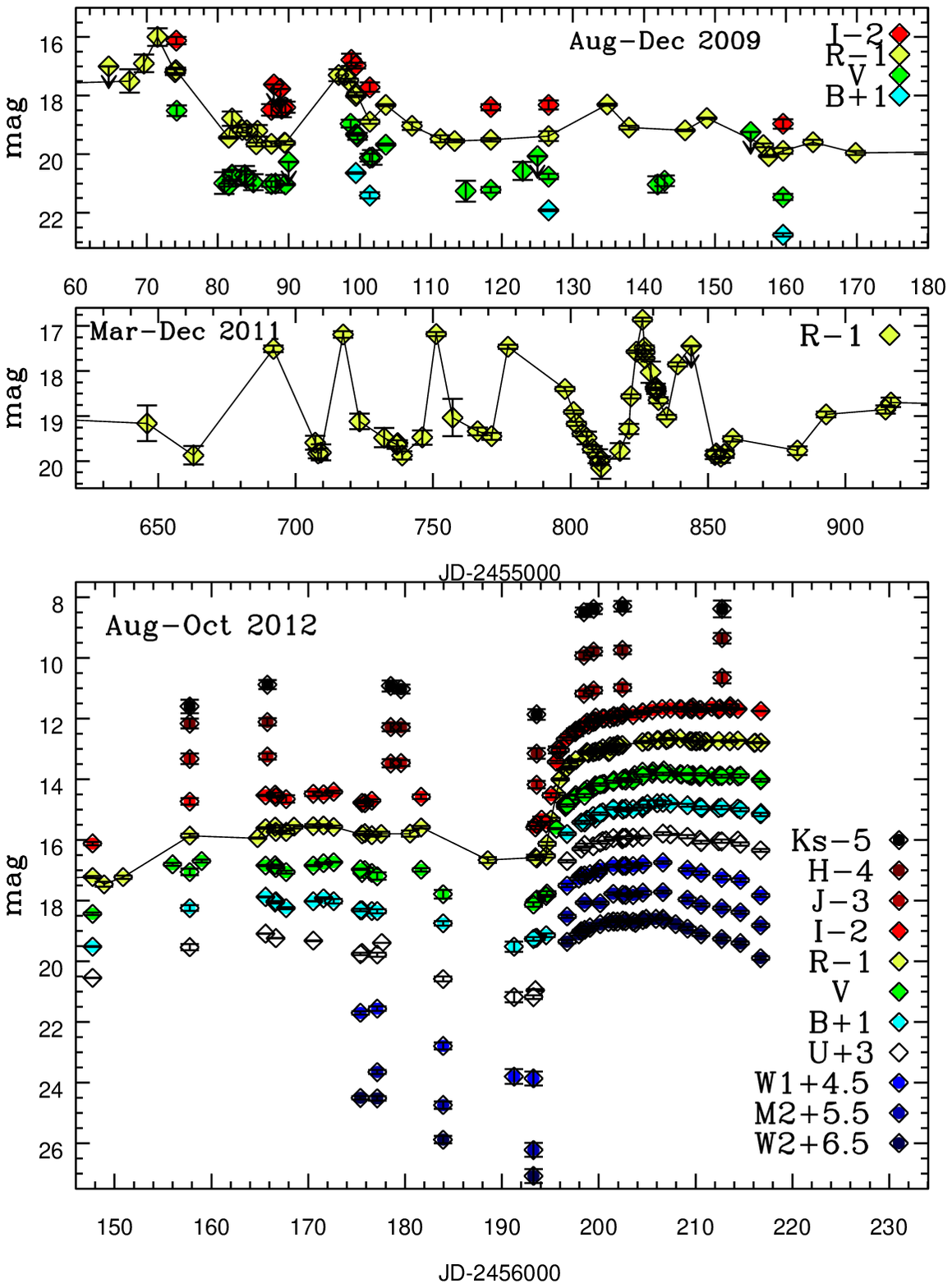}
\caption{{\bf Top:}  BVRI light curves of the impostor SN 2009ip during the first 3 months from the first
ever detection in 2009 \protect\citep{maza09}.
{\bf Middle:} R-band light curve of SN 2009ip during the period March to December 2011, showing erratic magnitude oscillations with $\Delta$R $\approx$ 3 mag. 
{\bf Bottom:} Ultra-violet/optical/near-infrared apparent light curves of the transient from August 8, 2012, 2 weeks before the 
publication of the announcement of a new re-brightening from \protect\citet{dra12}. Shifts of $\Delta$U = +0.27 ,  $\Delta$B = +0.018 and $\Delta$V = -0.042 have been applied to the u,b,v 
Swift/UVOT magnitudes
of SN 2009ip to match the U,B,V Johnson photometry. The shifts have been computed after a comparison of the magnitudes of the reference stars in the SN field in the two photometric systems.
\label{fig3}}
\end{figure*}

 The final photometry of the transient (Tables \ref{tab_opt}\footnote{We find excellent agreement with the 
CRTS and \citet{pri12b} photometry. The first version of \citet{mau12} posted to the arXiv archive showed a 
large disagreement with our photometry, being fainter by more than 2 mags in B band and 0.3 mags
in I band. The R-band data were in reasonable agreement to  the order of  a few hundredths  of a magnitude. 
Hence the average B-R color computed with the Mauerhan et al. photometry
was $\sim$ 2.5, vs. B-R $\sim$ 0.5-0.7 that is calculated with our data.  
 However, subsequent versions of the Mauerhan et al. paper (versions 2 and 3) show fair agreement with our photometry, with differences of few tenths 
of a magnitude.}, \ref{tab_swift} and \ref{tab_NIR}), along with  unpublished optical photometry of the comparison object NGC 3432-LBV1 (Table \ref{tab_00ch}).
The magnitudes of the reference stars in the field of NGC 7259 are listed in Table \ref{tab_seqstars}.

The R-band absolute light curve of SN 2009ip starting from August 2009
and spanning a period of more than 3 years is shown in Figure
\ref{fig2} along with that of a similar event, NGC
3432-LBV1 \citep{pasto10}, the debated transient (SN or impostor)
1961V \citep[photographic mags,][]{ber63,ber64,ber65,ber67} and the
revised visual light curve of the Giant Eruption of $\eta$ Carinae in
1842-1845 \citep[see][and references therein]{smi11}.  The same
distance modulus ($\mu$ = 31.55$\pm$ 0.15 mag) and interstellar extinction
(A$_R$ = 0.051) adopted by \citet{smi10} and \citet{mau12} for SN
2009ip have been used in the absolute R-band light curve of Figure
\ref{fig2}.  The erratic light curves of all these transients show
similar features. SN 2009ip experienced a few intense eruptive phases,
including those on August-September 2009 and from May to October 2011,
characterized by a sequence of sharp luminosity peaks followed by
rapid magnitude declines. 

 The detailed, multi-band light curve of SN 2009ip during different outbursts is shown in Figure  \ref{fig3},
with the 2009 event in the top panel (BVRI bands), the 2011 event in the middle panel (R band only)
and the 2012 outbursts in the bottom panel (from the ultra-violet SWIFT data to the NIR bands).

The 2009 and 2011 eruptive phases present the erratic evolution typical of on
LBV-type giant eruption, and is very similar to those observed in the
Giant Eruption of $\eta$ Carinae and in NGC 3432-LBV1.
Other re-brightenings were registered by
CRTS \citep[to magnitudes V$\sim$ 17 on Jul 15, 2010, and V$\sim$ 17.7 on Sep 29, 2010,][shown as cyan diamonds in Figure \ref{fig2}]{dra10}.
Older records (before August 2009) from the CRTS
archive\footnote{http://nesssi.cacr.caltech.edu/catalina/current.html}
and from \citet{smi10} have never registered the transient at a
magnitude brighter than about $\sim20.4$.  
These new data are more comprehensive, and reveal a recent variability history for SN 2009ip which is
more complex than one can infer from the schematic light curve representation of \citet{mau12}.

\begin{figure}
\includegraphics[angle=0,scale=.45]{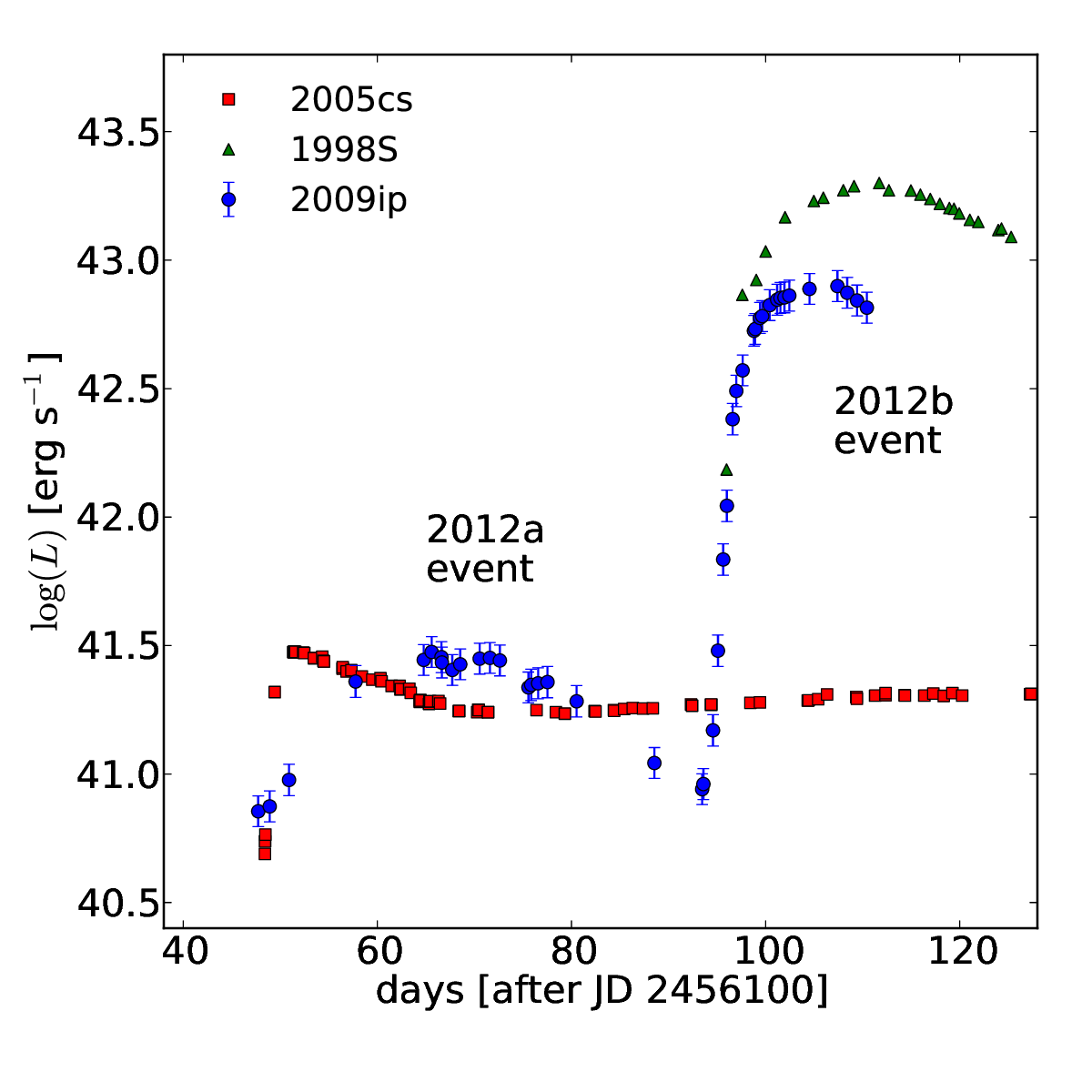}
\caption{Bolometric light curve of SN 2009ip from August to October 2012 (showing both the 2012a and 2012b events), 
compared with the bolometric light curves of the faint type IIP SN 2005cs
\citep{bro07,pasto06,pasto09} and the type IIn/IIL SN 1998S \citep{liu00,fas00,ger02,poz04}. The light curves of SNe 2005cs and 1998S are shown in an arbitrary temporal scale
to well match respectively the 2012a and 2012b eruptive events of SN 2009ip. \label{fig4} }
\end{figure}

During July-August 2012 a new re-brightening was announced by
\citet[][cyan diamonds in Figure \ref{fig2}]{dra12}. This event
was then followed by a strong unprecedented burst (starting around
September 23) which is about 30 times more luminous than the previous
oscillations.  This SN-like rise in luminosity will be extensively
discussed later in this paper. 

\subsection{Bolometric light curve} \label{bolo}

A pseudo-bolometric light curve was derived integrating the observed fluxes of SN 2009ip
from the ultra-violet to the NIR domains. In practice, for each epoch that had R-band observations available and
 for each band we derived the flux at the effective wavelength. When no observation in a given X filter was 
available for a specific epoch, the missing X-band photometric point was obtained through an interpolation of
the available data or, if necessary, by extrapolating the missing photometry assuming  a constant (R-X) color from 
the first/last available epoch.
The fluxes, corrected for the adopted extinction, provide the spectral energy distribution at the given 
phase, which is then integrated by the trapezoidal rule. The observed flux is finally converted in luminosity 
adopting the distance value mentioned above. We did not account for flux contribution outside the observed 
ultra-violet to NIR bands, and therefore this should be more properly quoted as a "quasi-bolometric" light curve.
Error estimates include the error in the photometry listed in Tables \ref{tab_opt}, \ref{tab_swift} and \ref{tab_NIR}, 
and the uncertainty in the distance modulus.

Figure \ref{fig4} shows the bolometric light curve of SN 2009ip from
the August 2012 re-brightening announced by \citet{dra12} to the
current epoch. It appears to show 2 distinct phases: a broader (and
fainter) earlier peak (that we will label as ``2012a event'' for
simplicity), that ends around September 23 and reaches a luminosity of
3 $\times$ 10$^{41}$ erg s$^{-1}$, and a fast-rising, higher luminosity
second peak (``2012b event'') with a maximum at about 8 $\times$
10$^{42}$ erg s$^{-1}$.  \citet{mau12} noted that the maximum
luminosity of the 2012a event is consistent with the luminosity of a
faint SN IIP \citep{pasto04}, although with a faster evolving light
curve.  Along with spectral similarities, this led \citet{mau12} to
suggest that the 2012a event was the actual core-collapse SN event 
of the LBV star.  We confirm that the bolometric luminosity of the
2012a event is similar to SN 2005cs, as one can note from Figure \ref{fig4}. 
The subsequent faster rise to the second peak  (the 2012b event)
presents an even tighter similarity with that of the type IIn/IIL SN
1998S. The 2012b event was proposed by \citet{mau12} to be 
due to strong SN ejecta-CSM interaction. We measure a 2 
week long rise-time, reaching a peak apparent magnitude of B = 13.80 (R = 13.65) on October 6, 2012
and then it declines in luminosity, more rapidly in the ultra-violet
and the blue optical bands.  We will see in Section \ref{discussion}
that the 2012a and 2012b sequence of events may have an alternative
explanation.

We also remark that none of the comparison objects in Figure
\ref{fig2} shows the regular, SN-like light curve that characterized
SN 2009ip during the 2012b event. This late photometric evolution
combined with the bright luminous peak (M$_R \approx$ -18) may support
the claim that {\it at least} during the 2012b event SN 2009ip has
finally exploded as a real supernova.  We note that the color of SN
2009ip at the light curve peak (on October 6, 2012) is U-V $\approx$
-1 mag, significantly bluer than that of the 2012a event at maximum
(U-V $\approx$ -0.5 mag).  At the pre-burst minimum of September 23,
the U-V color was instead significantly redder, i.e. $\approx$ 0 mag.

\subsection{Spectroscopy} \label{spectra}

\begin{deluxetable*}{ccccc}
\tabletypesize{\small}
\tablecaption{Log of observed spectra of SN 2009ip. \label{tab_spec}}
\tablehead{ \colhead{Date (dd/mm/yy)}           & \colhead{JD-2400000}      &
\colhead{Instrumental configuration}          & \colhead{Range (\AA)}  &
\colhead{Resolution (\AA)}        
}
\startdata
07/09/09 & 55081.57 & VLT(UT1)+FORS2+300V+300I & 3500-10350 & 10;9 \\
25/09/09 & 55099.58 & VLT(UT1)+FORS2+300V+300I & 3250-10000 & 10;9 \\ 
29/09/09 & 55103.66 & VLT(UT1)+FORS2+300V+300I & 3500-10030 & 10;9 \\
22/10/09 & 55126.7 & NTT+EFOSC2+gm11+gm16 & 3530-9440 & 14;12 \\
24/11/09 & 55159.58 & NTT+EFOSC2+gm11 & 3350-7470 & 14 \\
06/10/10 & 55475.60 & NTT+EFOSC2+gm11+gm16 & 3360-9540 & 21;20 \\
02/09/11 & 55807.48 & VLT(UT2)+XShooter & 3030-10400 & 1.0;0.8 \\
24/09/11 & 55828.64 & VLT(UT2)+XShooter & 3150-22900 & 1.0;0.8;2.8 \\
08/08/12 & 56148.93 & NTT+EFOSC2+gm11 & 3360-7470 & 14 \\
18/08/12 & 56157.76 & VLT(UT2)+XShooter & 3100-24790 & 1.0;0.8;2.8 \\ 
25/08/12 & 56164.77 & NTT+EFOSC2+gm11 & 3390-7450 & 14 \\
26/08/12 & 56165.58 & NTT+EFOSC2+gm11 & 3360-7450 & 14 \\
26/08/12 & 56165.78 & NTT+SOFI+GB & 9370-16440 & 27 \\
27/08/12 & 56166.64 & NTT+EFOSC2+gm13 & 3650-9250 & 18 \\
29/08/12 & 56168.50 & TNG+Dolores+LRB+LRR & 3170-9800 & 10.5;9.5\\
30/08/12 & 56169.54 & TNG+Dolores+LRB & 3280-8080 & 10.5 \\
31/08/12 & 56170.50 & TNG+Dolores+LRB+LRR & 3280-9300 & 14;13 \\
05/09/12 & 56175.55 & NOT+ALFOSC+gm4 &  3350-9070 & 18 \\
10/09/12 & 56180.56 & NTT+EFOSC2+gm11+gm16 & 3360-10040 & 22;20 \\
18/09/12 & 56188.55 & NTT+EFOSC2+gm11 & 3360-7450 & 14 \\
21/09/12 & 56192.43 & WHT+ISIS+R300B+R158R & 3200-9250 & 4.3;7.2 \\
22/09/12 & 56192.52 & TNG+Dolores+LRB & 3320-8080 & 14 \\
23/09/12 & 56193.50 & WHT+ISIS+R300B+R158R & 3180-9490 & 8.6;14 \\
23/09/12 & 56193.53 & NTT+SOFI+GB+GR & 9370-25200 & 27;30 \\
28/09/12 & 56198.80 & GN+GMOS+R400 +B600& 3400-9130 & 2.7,3.7 \\ 
04/10/12 & 56205.40 & TNG+Dolores+LRB & 3320-8090 & 10 \\
\enddata
\end{deluxetable*}

\begin{figure*}
\includegraphics[angle=0,scale=.83]{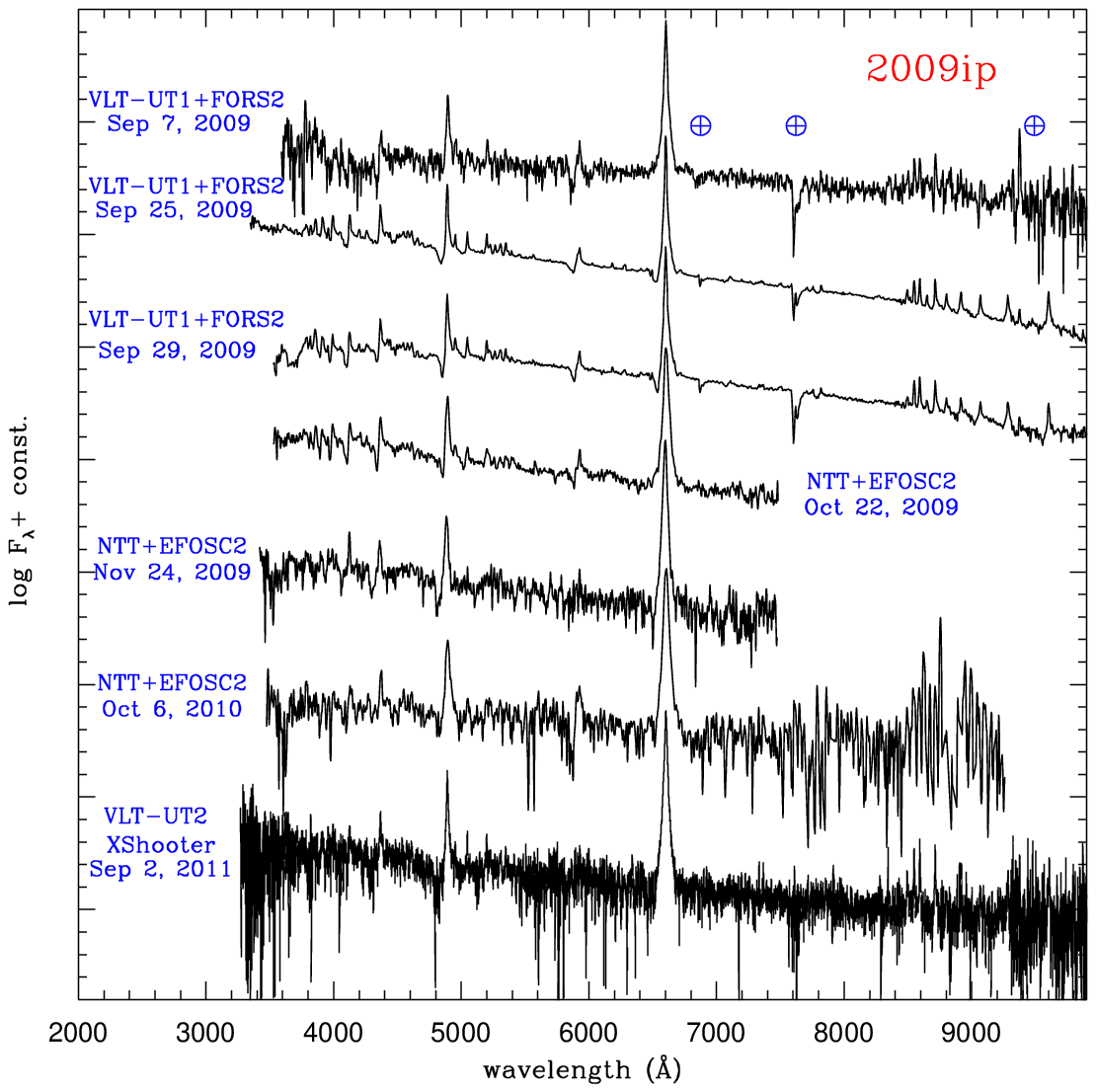}
\caption{Sequence of spectra of the LBV in NGC 7259, obtained from September 2009 to September 2011.
All spectra are in the host galaxy wavelength frame. The symbols ``$\oplus$'' mark the positions of the strongest telluric absorption bands.
\label{fig5}}
\end{figure*}

\begin{figure*}
\includegraphics[angle=0,scale=.83]{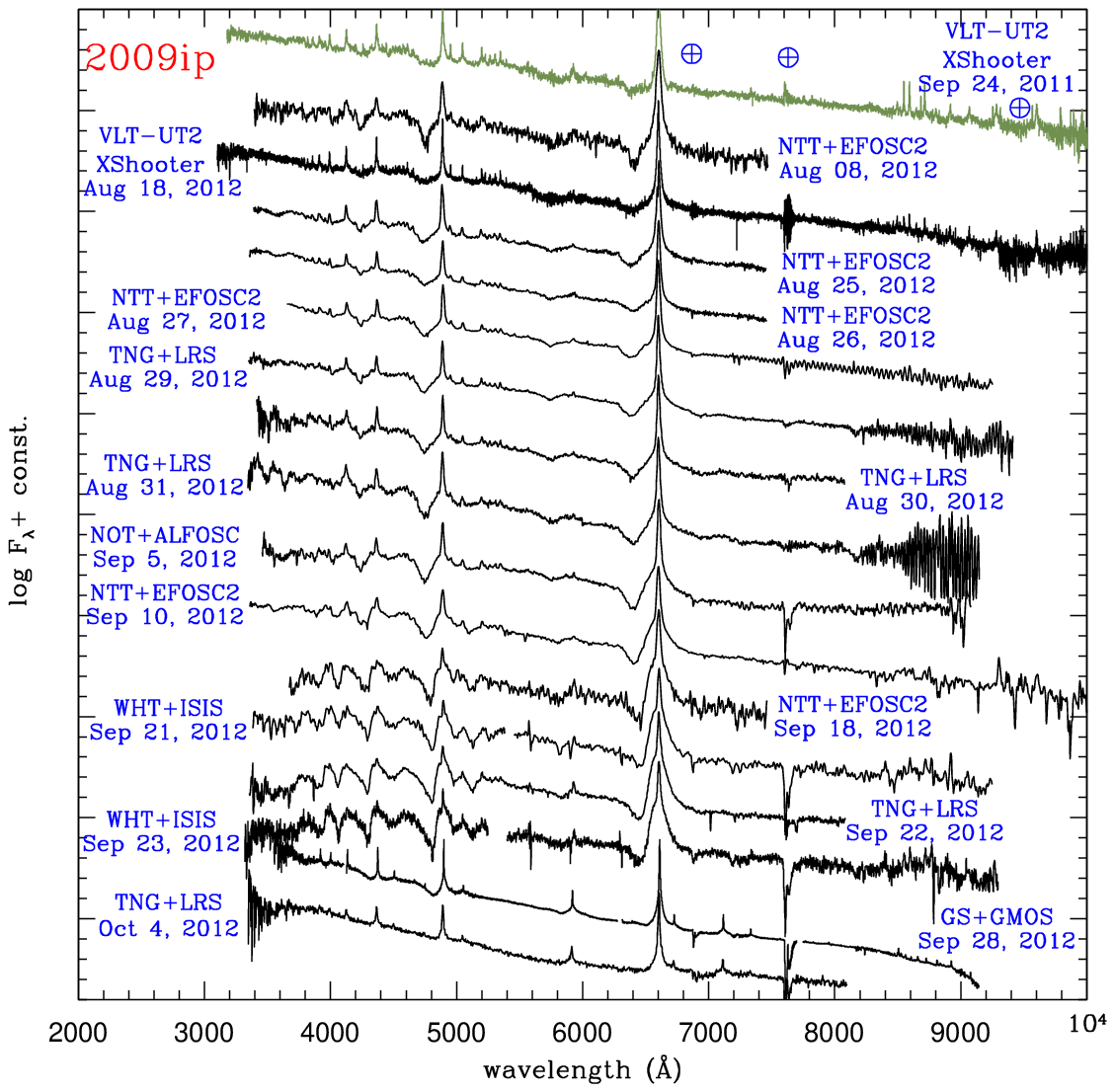}
\caption{Sequence of spectra obtained between August and September 2012, including those of the putative SN explosion. A higher resolution  XShooter
spectrum obtained on September 24, 2011, i.e. before the 2012 re-brightening, is also shown at the top of the sequence (green color).
All spectra are all in the host galaxy wavelength frame. The symbols ``$\oplus$'' mark the positions of the strongest telluric absorption bands.
\label{fig6}}
\end{figure*}

\begin{figure*}
\includegraphics[angle=0,scale=.85]{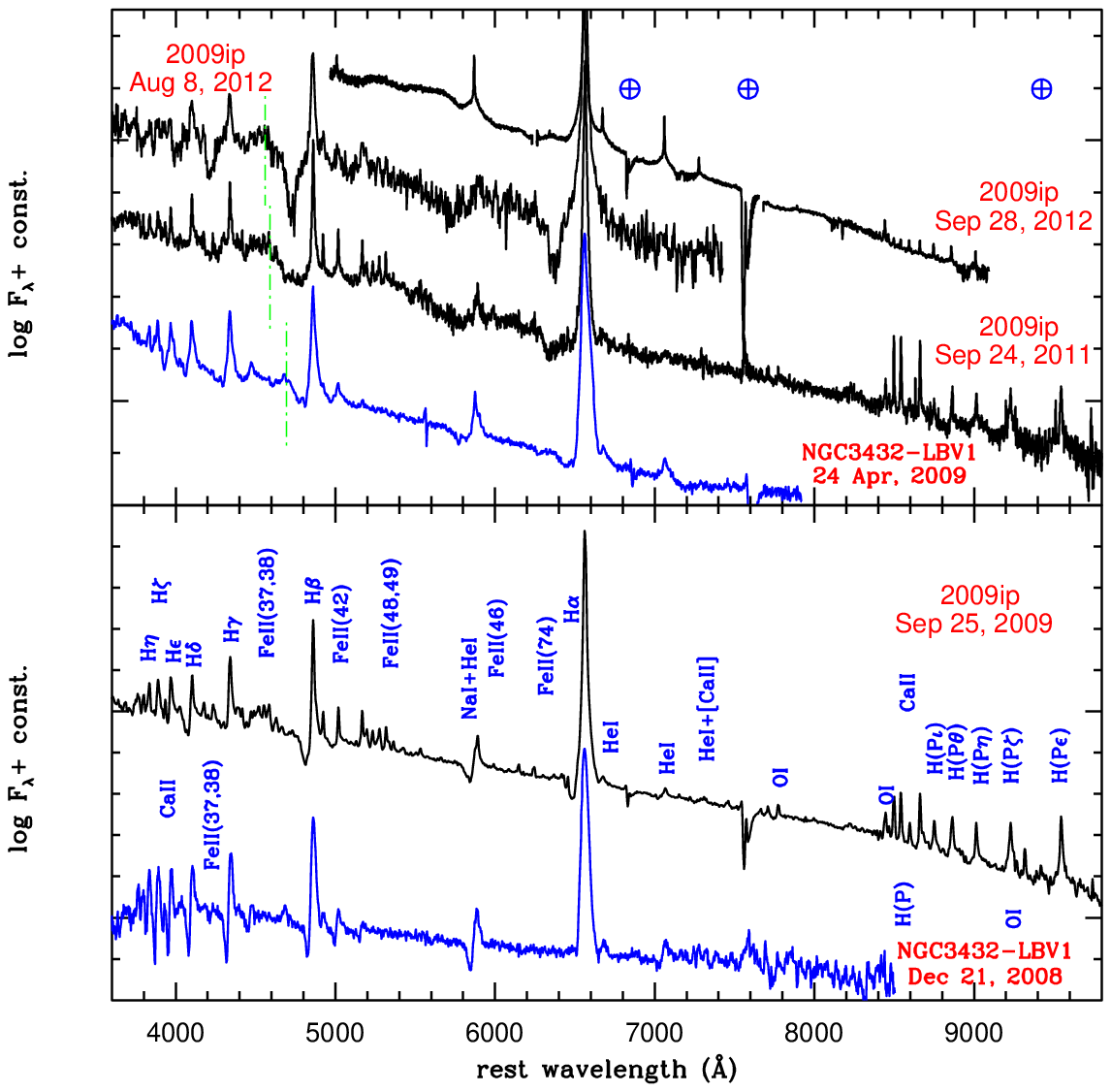}
\caption{{\bf Top:} comparison of spectra of SN 2009ip at 3 representative epochs (24 September 2011, and during the 2012a and 2012b events) with a spectrum in outburst of NGC3432-LBV1. 
The vertical dashed green lines mark the position of the highest velocity edges of the H$\beta$ components in the 2 objects. {\bf Bottom:}
line identification in the optical spectrum of SN 2009ip obtained on September 25, 2009 (VLT+XShooter), 
and comparison with a spectrum of NGC 3432-LBV1 in outburst. The symbols ``$\oplus$'' mark the positions of the strongest telluric absorption bands in the spectra of 
SN 2009ip. \label{fig7} }
\end{figure*}

\begin{figure*}
\centering
\includegraphics[angle=0,scale=.65]{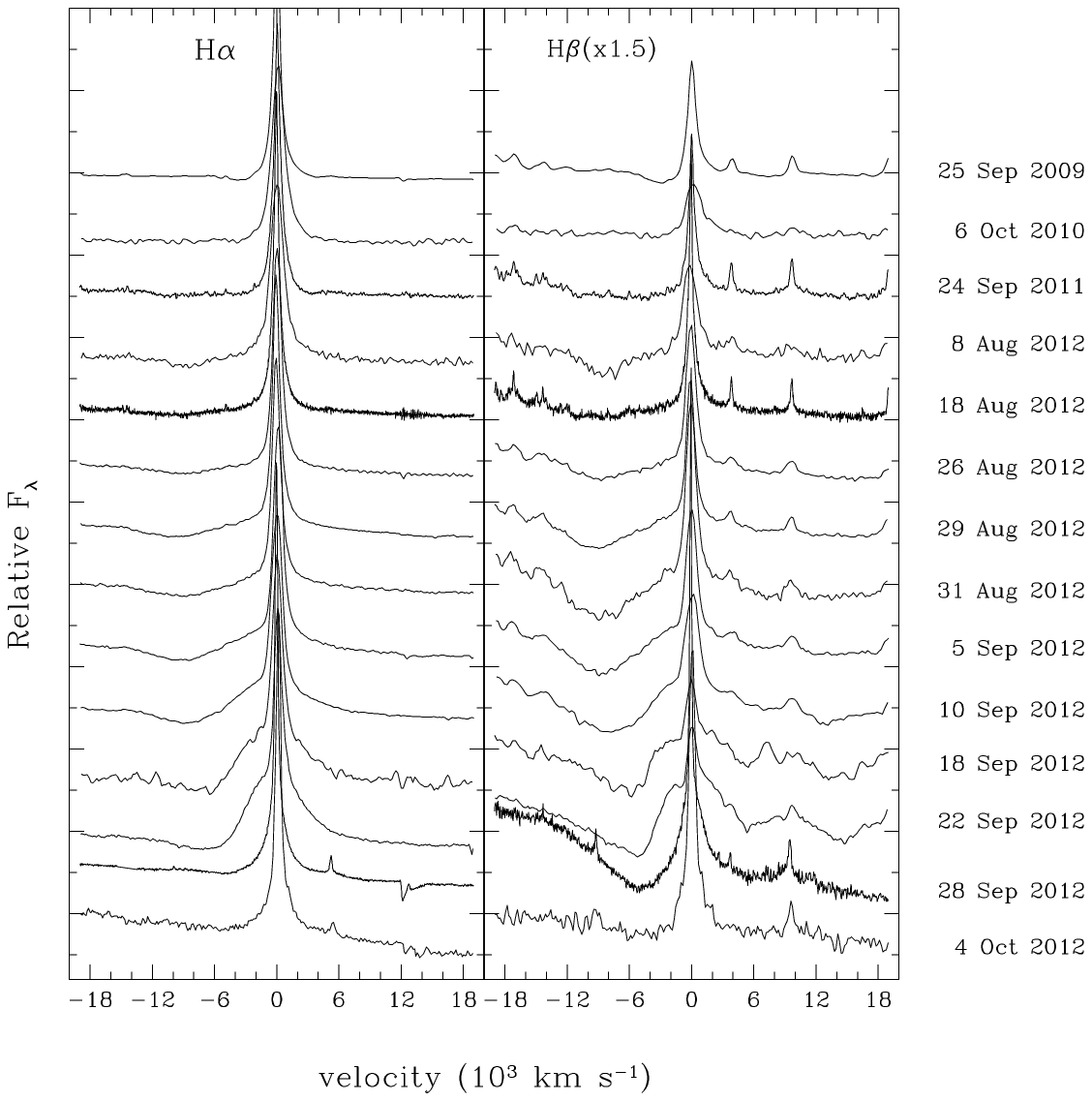}
\caption{Blow-up of the regions of H$\alpha$ (left) and H$\beta$ (right) for a selected sub-sample of SN 2009ip spectra. The velocities in abscissa are in units of 10$^3$ km s$^{-1}$. \label{fig8} }
\end{figure*}

\begin{figure*}
\centering
\includegraphics[angle=0,scale=.7]{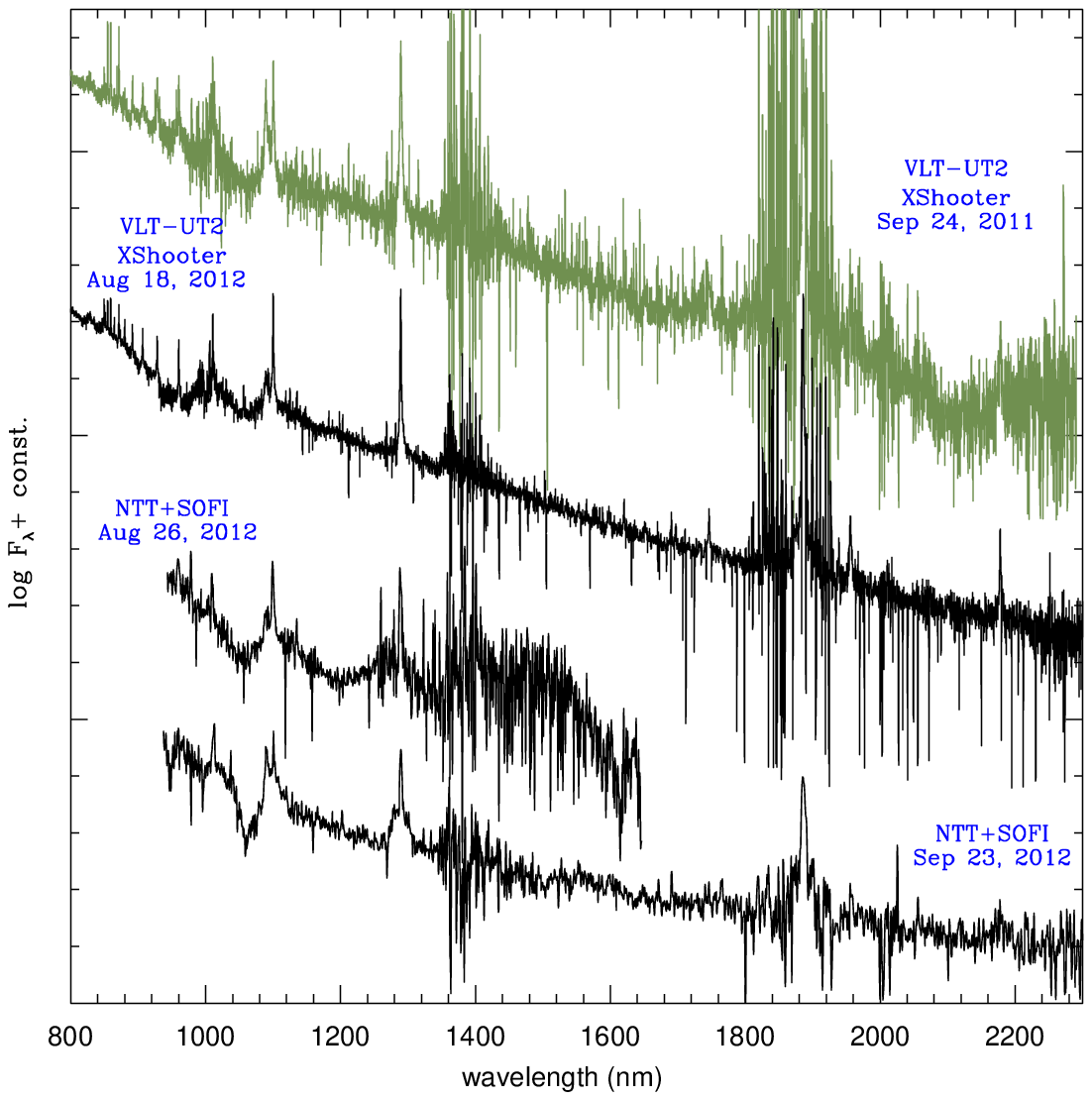}
\caption{Sequence of near-infrared spectra of SN 2009ip obtained from August to September 2012. The XShooter spectrum of September 24, 2011
is also shown in green. \label{fig9} }
\end{figure*}

Optical and near-infrared spectra of SN 2009ip (Figures \ref{fig5}, \ref{fig6}, \ref{fig7}, \ref{fig8} and \ref{fig9}) 
were collected using the 8.2-m Very Large Telescope (VLT) UT1 (+ FORS) and UT2 (+XShooter) at the Cerro Paranal Observatory (ESO Chile), 
the 3.58-m ESO-NTT (+ EFOSC2 and SOFI) at the La Silla Observatory (ESO Chile), the 8.2-m Gemini North Telescope (with GMOS) in Cerro Pach\'on (Chile),
the 3.58-m Telescopio Nazionale Galileo (TNG, equipped with LRS), the 4.2-m  William Herschel Telescope (WHT, with ISIS) and the 2.56-m 
Nordic Optical Telescope (+ ALFOSC) located in La Palma (Canary Islands, Spain). Basic information on the spectra collected during the observational campaign
of SN 2009ip is reported in Table \ref{tab_spec}.

The spectroscopy data reduction steps were performed using IRAF tasks. The pre-reduction process (i.e. overscan and bias corrections, flat-fielding and trimming) 
is the same as described for the photometry images. In addition, for the IR spectra, the contribution of the night sky emission was removed by subtracting 
from each other two consecutive exposures taken with the source in different positions along the slit. The optimal extraction of 1-dimensional spectra 
allowed us to remove the flux contamination of the night sky (for the optical spectra) and other background sources.
The spectra were then wavelength calibrated using reference spectra of arc lamps, and calibrated in flux using sensitivity curves 
obtained through spectra of spectro-photometric standards. The consistency of the spectroscopic flux calibration was finally 
checked using the available SN photometry and, when discrepant, the spectral fluxes were rescaled. 
Telluric standards were  used to correct the NIR spectra for the effects of the broad atmospheric absorption bands.

The spectra relative to the 2009 outburst reported in Figure \ref{fig5} are all dominated by prominent Balmer lines with a complex profile. 
The weak absorption features indicate that the bulk of the ejected material is moving with a velocity of 2900 $\pm$ 700 km s$^{-1}$, but the blue 
edge of the isolated H$\beta$ absorption suggests the presence of fast-moving material which is expanding at a velocity of about 5000-6000 km s$^{-1}$. 
The H$\alpha$ emission component in September 2009 has a Lorentzian profile with a FWHM velocity of about 700-800 km s$^{-1}$, which
increases to about 1100-1200 km s$^{-1}$ during the period October-November, 2009 (when the object was receding to a more quiescent stage).

After a further outburst (September 2010) reported by \citet{dra10}, a spectrum obtained on October 6, 2010 shows SN 2009ip at a similar stage as the November 24, 2009 
spectrum, i.e. with the star again quiescent. The FWHM velocity of the Lorentzian H$\alpha$ component in this phase is still around 1300 km s$^{-1}$.
The September 2, 2011 VLT spectrum reported at the bottom of Figure \ref{fig5} shows SN 2009ip to be back to a dormant stage, and the FWHM velocity of the 
Lorentzian H$\alpha$ component is about 940 km s$^{-1}$.

Figure \ref{fig6} shows the spectra of the transient during the period
August-October 2012, compared with a VLT spectrum obtained on
September 24, 2011 (green line), during another outburst episode. In
the September 24, 2011 spectrum, the FWHM velocity of H$\alpha$, which
still has a Lorentzian profile, has slightly decreased to around 790
km s$^{-1}$, and other Balmer lines clearly show very broad absorption
components, with a blue edge that indicates that there is material
moving with a velocity as high as 12500 km s$^{-1}$ {\sl already} at
this epoch (see also Figure \ref{fig7}).  This is the highest velocity 
outflow that has been detected in an LBV-like eruption of any sort 
and indicates that high velocities are observed without core-collapse
or the catastrophic destruction of the star. This has important
consequences for the interpretation of high velocity ejecta as
evidence for the core-collapse mechanism in the 2012a event. 
\footnote{We are confident that the broad line absorptions observed in the September 2011 XShooter spectrum
are intrinsic to the object and not artifacts due to the instrumental effects. 
In fact, the major problem that could affect the line shapes (in particular the absorptions) 
could be the blaze function being not properly corrected at the edges of the spectral orders. 
However the resulting patterns would affect a wider wavelengths range ($\sim$1000 \AA) and would have a smoother 
effect rather than mimic a single line absorption profile \citep[see, e.g., the 2010 March 23 spectrum of SN 2010bh in][]{buf12}.}

We subsequently obtained an NTT
spectrum on August 8, 2012 \citep[JD = 2456148.91, i.e. 10 days before
the new outburst - the 2012a event - was announced by][]{dra12}.
The broad absorption features were present also at this epoch, and
indeed were stronger than in the September 24, 2011 spectrum (Figure
\ref{fig6}).  The minimum of the broad absorption components of the
Balmer lines has a core at 8600 $\pm$ 400 km s$^{-1}$, with a blue
wing extending up to 14000 km s$^{-1}$, while the Lorentzian emission
survives at a FWHM velocity of about 1380 km s$^{-1}$.  The presence
of these components was observed in September 15 and 16, 2012 spectra
by \citet[][]{smi12a}, and this was the critical measurement that led
the  authors to propose that the LBV had exploded as a core-collapse SN, i.e. that
the 2012a event was due to stellar core-collapse and an explosion with
fairly  low kinetic energy like SN 2005cs. 

Our spectra collected between August 18 and September 5, 2012 show little evolution: the H features show prominent P-Cygni profiles,
with deep minima at 8000-9000  km s$^{-1}$ and edges possibly extending to 14000-15000 km s$^{-1}$. The H$\alpha$ narrow emission component still has a FWHM velocity of 800 $\pm$ 100
km s$^{-1}$, while the highest resolution spectra allow us to measure the FWHM velocity from the clearly detected narrow Fe II emissions (multiplet 42) to be about 240 $\pm$ 20  km s$^{-1}$.

As highlighted by \citet{mau12}, the spectra from September 10 to 23
(2012a event) do closely resemble those of type II SNe (the similarity
with early spectra of the under-luminous type IIP SN 2005cs shown in
their Figure 2 is remarkable). Both H and Fe II lines now show broad
P-Cygni profiles with a prominent broad emission component. 
However we now present spectra of the 2012a event covering a
period from August 8, 2012 to September 23, 2012 (47 days), and we do not observe the
typical evolution of a type II SN over this period. In particular,
15-20 days after explosion, type II-P SNe develop the strong, broad 
near-infrared Ca II triplet feature \citep{pasto06}, but we don't
observe this for the 2012a event. The cores of the absorptions of the Balmer
features indicate expansion velocity of the ejected material of
$\sim$ 5000-6000 km s$^{-1}$ (4200 $\pm$ 500 km s$^{-1}$ from the Fe
II lines), but the blue edge of the wings still reach to much higher velocities
(about 13800 km s$^{-1}$). Figure \ref{fig7} (top) shows
a comparison of SN 2009ip at 3 representative epochs (September 24,
2011; August 8 and September 28, 2012) with a spectrum of NGC
3432-LBV1 in outburst (April 24 2009). The high velocity 
P-Cygni absorption (in the Balmer lines) is certainly stronger in the 2012a event than 
we observed in 2011 and in NGC3432-LBV1 in outburst, but we illustrate
here that the detection of high velocity gas is not only 
restricted to core-collapse SNe. Similar high velocity edges are clearly detected in SN
2009ip in 2011 (13800 km s$^{-1}$) and in NGC3432-LBV1 \citep[$\sim$9000 km
s$^{-1}$,][]{pasto10}. A blow up of the H$\alpha$ and H$\beta$ regions of
a few selected of spectra of SN 2009ip is shown in Figure \ref{fig8}, supporting our claim that broad
absorption features -though fainter- were detected even before August 2012. 
We will discuss the implications of this in Section \ref{discussion}.

These broad absorptions disappear at the
time of the 2012b event, in September 28 and October 4 spectra
(Figures \ref{fig6} $\&$ \ref{fig7}, top), when the luminosity of SN
2009ip reaches the unprecedented maximum. At these times, the spectra
are very similar to those of many type IIn SNe \citep[e.g. SN
1999el,][]{dica02}, with the H lines presenting a narrow emission
component with a FWHM velocity of about 290 km s$^{-1}$ and very broad
wings ($\sim$3600 km s$^{-1}$). Similar velocities are measured in the
He I lines, which are now more prominent than in past spectra, whilst
the Fe II lines are no longer visible.  

The spectrum of SN 2009ip obtained on 25 September 2009 with VLT-UT1
equipped with FORS2 has a very high signal-to-noise ratio. This gives
us the opportunity to identify the most important lines in the
spectrum of SN 2009ip (Figure \ref{fig7}, bottom). The spectrum is
dominated by strong Balmer and Paschen lines of H, showing weak and
narrow (2850 $\pm$ 490 km s$^{-1}$) P-Cygni profiles. Weak He I lines
(being the 5876\AA~ feature blended with Na I D 5890-5896 \AA) and a
number of Fe II multiplet lines are also detected. We note that in the
September 28, 2012 spectrum of SN 2009ip (during the 2012b outburst,
Figure \ref{fig7}), the spectral properties are quite similar to those
observed in the afore-mentioned VLT spectrum, although the detection
of Fe II lines is not obvious. Most of these lines are also visible in
the spectrum of the impostor NGC3432-LBV1 shown as a comparison, with
quite similar velocity of the narrow components ($\leq$ 650 km
s$^{-1}$). Narrow O I and Ca II lines are relatively prominent in SN
2009ip, while they were not unambiguously detected in NGC 3432-LBV1
(although this might be due to the lower signal-to-noise spectrum).

A sequence of near-infrared spectra of SN 2009ip is shown in Figure \ref{fig9}. The continuum is always quite blue in these
spectra. The strongest lines are detected as broad features with P-Cygni profiles, and narrower emissions 
superimposed to the broad components. The broad P-Cygni components become more evident with time and in the September 23, 2012 spectrum (at the time of the onset of the 2012b eruption) they dominate over the narrow lines.
We identify Br $\gamma$ at 2165 nm, Pa $\alpha$ (that is barely visible in the middle of the telluric absorption around 1875 nm), 
Pa $\beta$ at 1282 nm and Pa $\gamma$ at 1094 nm, blended with He I 1083 nm.
The September 23, 2012 spectrum, in particular, shows a broad Pa $\beta$ with FWHM velocity of about 6200 km s$^{-1}$ and a prominent blue-shifted absorption of Pa $\gamma$ + He I 1083 nm with an
expansion velocity of about 10000 km s$^{-1}$, as obtained from the position of the broad absorption minimum. The narrow He I 1083 nm line, which was marginally detectable in previous spectra, is now clearly
visible, and is well separated from Pa $\gamma$. 
The narrow Paschen lines have Lorentzian profiles with a FWHM velocity of about 400 km s$^{-1}$, whilst the narrow He I $\lambda$ 1083 nm appears to be slightly broader ($\sim$800-1000 km s$^{-1}$) and with a 
roughly Gaussian profile.

\section{Real Supernova or Supernova Impostor?} \label{discussion}

SN 2009ip is a remarkable object for a number of reasons: {\bf i)} it
experienced a series of energetic outbursts since 2009, when the
transient reached absolute peak magnitudes between -14 and -15; {\bf
  ii)} the spectral features reveal the presence of ejected material
at very high velocities (several $\times$ 10$^{3}$ km s$^{-1}$); {\bf
  iii)} the progenitor star was observed to be extremely luminous in
quiescence (M$_V$ = -10.0 $\pm$ 0.3) and was proposed to be a massive
LBV \citep[$>$ 60 M$_\odot$,][]{fol11,smi10}; {\bf iv)} finally, in
September 2012 the star displayed a further, exceptionally luminous
outburst \citep[the 2012b event, with M$_V$ $\approx$ -18,][and
references therein]{mau12}, suggesting that the LBV may have 
experienced a core-collapse SN explosion. The luminosity during
  that event, and its similarity to SNe IIn spectra are possibly the
  strongest indicators that a core-collapse SN has occurred, more so
  than the broad lines of the spectra during the 2012a pre-cursor event.

The complex, erratic 2009-2012 light curve of SN 2009ip (Section \ref{photo}) indicates that the LBV entered a very active variability phase resembling those of the unusual
NGC 3432-LBV1 or $\eta$ Carinae during the Giant Eruption of the 19th century. In the case of NGC 3432-LBV1, multiple eruptions on short time scales (about 200-220 days) have been proposed 
to be the result of violent pulses of a very massive star (possibly
via the pulsational pair-instability mechanism) that is approaching
the end of its life, presumably with the core-collapse. Alternatively, the pulses 
may be regulated by the passage of a companion star to the periastron \citep{pasto10}\footnote{The presence of a companion was proposed to explain the modulated, quasi-periodic light curve of NGC 3432-LBV1 \citep{pasto10}.}.

 The presence of very fast material ($\sim13000$ km s$^{-1}$) in
  SN 2009ip {\it already almost 1 year before the putative SN
    explosion (i.e. in the 24th September 2011 spectrum)}, and also in
  NGC 3432-LBV1 ($\sim9000$ km s$^{-1}$) suggests that these LBV
  related eruptions could quite feasibly be linked with the 2012a
  event.  The highest velocity  in the Homunculus 
Nebula surrounding $\eta$ Carinae  reaches $3500-6000$ km s$^{-1}$
\citep{smi08}.
Typical LBV eruptions are discussed in terms of extreme stellar winds
driven by the super-Eddington luminosity of the star.  However, these
winds are expected to have velocities of the order of a few $\times$
10$^{2}$ km s$^{-1}$ \citep[e.g.][]{smi08}.
The detection of this high-velocity gas in some LBV outbursts (including
the afore-mentioned events) suggests that these episodes probably
originate in explosions deeper in the star, perhaps in the core. 
These release an energy that may compete 
with those of weak SNe \citep[e.g. faint SNe IIP, such as SN 1999br,][]{pasto04},
producing a blast wave that allows the star to expel massive portions of the envelope \citep{smi08}. All of this is expected to produce transients  that closely mimic the energy and the overall
properties of a real SN exploding in a dense CSM (type IIn).

\subsection{No core-collapse SN during the 2012a event?} \label{noCCSN}

One of the most remarkable findings inferred from the analysis of the August and early September spectra of SN 2009ip (during the 2012a event)
is that the bulk of the ejected  material has extremely high expansion velocities \citep[8000-9000 km s$^{-1}$, with edges extending up to 14000 km s$^{-1}$, see Section \ref{spectra}, and][]{mau12}. 
This, and the striking similarity between the early September spectra of SN 2009ip and those of the weak type IIP SN 2005cs \citep{pasto06,pasto09} led \citet{mau12} to conclude that SN 2009ip had 
likely exploded as a faint, $^{56}$Ni-poor core-collapse SN during the August re-brightening episode. 
The fact that we can observe features from the SN ejecta inside an extended and dense CSM is explained with a non homogeneous, possibly clumpy distribution of the material lost
by the LBV in pulsations preceding the explosion.
 While this is plausible, we would caution that the detection of
  high-velocity ejecta cannot be regarded as a conclusive proof,
  because very high velocity material was also observed in NGC
  3432-LBV1 
\citep[][where the broad wing of the H$\beta$ absorption extended
 to 9000 km s$^{-1}$]{pasto10}, during an eruption of a known SN
 impostor. 

The core-collapse SN scenario proposed by \citet{mau12} is questionable,
since there is a number of observables that require a rather ad-hoc
combination of events: {\bf i)} the high-velocity absorption wings
measured in the spectra obtained after the announcement of the 2012a
outburst episode \citep{dra12} are actually similar to those we have
seen in the September 24, 2011 and August 8, 2012 spectra, which
raises the question whether and in case when the SN explosion
occurred; {\bf ii)} the peak absolute magnitude (M$_R \sim$ -15) and
the evolutionary timescales of the 2012a event are consistent with
those of previous eruptive episodes (in particular the 2009 event, see
Figure \ref{fig3}); finally {\bf iii)} it is not trivial to explain
how an extremely massive LBV (M$>$60 M$_\odot$, likely with M$_{ZAMS}
\geq$ 90-100 M$_\odot$) can explode as a weak type II SN: we may need
to invoke sub-sequent eruptions to explain the events before
July-August 2012, and subsequently a fall-back core-collapse SN with
formation of a black hole.

In the  \citet{mau12} interpretation, the 
2012b event is fairly simply explained as core-collapse SN ejecta-CSM
interaction.  However it is also plausible that the 2012a event was
an eruptive phase, and the 2012b luminosity comes from the actual
core-collapse, similar to what is assumed to occur in IIn SNe,
or from the collision of material ejected in the previous eruption
with pre-existing CSM, or even from the illumination of the inner parts
of a dense circumstellar disk by faster ejected material \citep[as proposed by][]{lev12}.

 An alternative explanation for the nature of the 2012a outburst has been offered
by \citet{sok12}, who noted some similarity between the 2012a+2012b light curve 
of SN 2009ip with that of the unusual eruptive variable V838Mon \citep{tyl05,tyl06}, and 
proposed that the ejection of fast material following the merging of two massive 
stars might explain the 2012a event, whilst the subsequent collision of this fast 
material with pre-existing CSM would produce the 2012b event.

\subsection{SN 2009ip, a pulsational pair-instability event}

The detection of high velocity ejecta (12500\,km s$^{-1}$) on September 24, 2011 indicates that the star has managed to eject material at velocities that we would 
normally associate with a SN explosion. It is very unlikely that the core collapsed at this point (see Section \ref{noCCSN}),
which implies that the high velocity material has been ejected in the 2012a event without invoking a core-collapse SN explosion. 
What triggers these ejections is still unclear, but the very high progenitor mass \citep{smi10,fol11} indicates that the events may be signatures of pulsational 
pair-instability \citep{rak67,bar67,woo07}. The pulsational pair-instability scenario discussed by \citet{woo07} is applicable for stars with main-sequence masses 
in the range 95-130 M$_\odot$. This is apparently above the mass proposed for the precursor of SN 2009ip.  We note, however, that the absolute magnitude of the LBV 
progenitor of SN 2009ip \citep[see e.g. Figure 3 in][]{fol11} is also consistent with evolutionary tracks of stellar masses that are much higher than
60 M$_\odot$, and so should be regarded as a lower mass limit.

The \citet{woo07} model of a pulsational pair-instability SN suggests that colliding shells of material can dissipate most of the relative kinetic energy as radiation.
One solar mass of material moving at 8000\,km s$^{-1}$ has a kinetic energy of more than $10^{50}$\,erg, enough to power the measured bolometric light curve  of the 2012b event shown in Figure \ref{fig4}. 
As SN 2009ip has experienced multiple mass ejections, perhaps even more than those we have detected due to possible gaps in the observational coverage (Figure \ref{fig2}), 
it is plausible there are shells, or clumps of slower moving gas that will slow the fast ejecta of 2009ip during the 2012a episode.

As discussed in \citet{mau12}, there are no known line-driven wind mechanism or continuum driven wind mechanism for driving material off the stellar surface at the high 
velocities observed. The energy to provide $\gtrsim 10^{50}$\,erg per solar mass ejected must presumably come from a core-related event.  

There is also some consistency in the velocity of the material ejected during the 2012a event and the radius of the emitting region in the 2012b episode.
The 2012a event lasts approximately 50 days, during which the bulk of material starts at 8000-9000\,km s$^{-1}$ on 5 September 2012, slowing to 5000-6000\,km s$^{-1}$ 
after about 10 days. The fast ejecta likely travelled around $5\times10^4$\rsun, before impacting on a surrounding shell and causing the dramatic increase in luminosity 
in the rise to the 2012b light curve peak. If the kinetic energy of the shell is then converted into radiative energy, one
would expect that an emitting sphere of radius $5\times10^4$\rsun~at a black-body temperature of around 10000 K would
emit at $L \simeq$ a few $\times 10^{43}$\,ergs$^{-1}$. This crude luminosity estimate is of the same order of magnitude to that we see in Figure \ref{fig4}.

The pulsational pair-instability SN model requires a star of initial mass to be in the range 95-130 M$_\odot$. The standard mass-loss prescriptions for such massive stars has to be relaxed 
so that in the final stages the star should retain enough mass to enhance the core temperature to cause the pair-instability. The progenitor has been estimated to have
more than 60 M$_\odot$, implying that has retained most of its envelope. This is supported by the evidence that broad hydrogen features are detected in all the ejection episodes 
\citep[][see also Section \ref{spectra}]{smi10,fol11}.

An interesting measurement would be the metallicity at the distance of SN 2009ip from the host galaxy nucleus (about 4\,kpc) to determine if the environment is significantly metal poor. 
At the current stage, only a statistical approach is possible to grossly estimate the local oxygen abundance. Adopting the host galaxy distance and reddening of \citet{smi10}, 
the host galaxy has an absolute B-band magnitude of -17.9. Following \citet{pil04}, the characteristic (at R = 0.4R$_{25}$) oxygen abundance  of NGC 7259 would be 12 + log(O/H) = 8.34,
 which gives 12 + log(O/H) = 8.07 at the SN position, clearly sub-solar. Although this method may provide -in the best case- 
 only a rough estimate of the oxygen abundance, a sub-solar metallicity is the natural expectation from the modest host galaxy brightness and the peripheral location
of the transient.

As a consequence, a pulsational pair-instability scenario  may provide a plausible explanation for the 2012 events, without necessarily invoking the core-collapse of the star. According to this,
the 2012a event may have been a pulsational pair-instability eruption followed by collisions of these ejecta with pre-existing CSM. The late September - early October spectra of SN 2009ip, again dominated by narrow lines 
with Lorentzian profiles, indicate that the high-velocity  material is covered by electron scattering in a high-density interaction shell \citep{mau12,chu04}. We note that, as mentioned in Section \ref{intro}, 
there is a relatively weak X-ray emission (L$_X \sim$ 4$\times$10$^{39}$ erg s$^{-1}$ at maximum)\footnote{Strongly interacting type IIn SNe have been observed to reach peak X-ray luminosities 
L$_X \approx$ 10$^{41}-$10$^{42}$ erg s$^{-1}$ \citep[][and references therein]{dwa12}. As a comparison, stripped-envelope core-collapse SNe have L$_X$ in the range 10$^{38}-$10$^{40}-$ erg s$^{-1}$, 
whilst for SNe IIP, usually  L$_X \ll$ 10$^{38}$ is found. We note that the X-ray emission of SN 2009ip is close
to that of SN 2011ht \citep{rom12}, whose nature -real SN or SN impostor- has not been firmly established.} and no radio detection of SN 2009ip. Although this would not support strong ejecta-CSM interaction, it does not necessarily rule it out.
According to the pulsational pair-instability scenario, the star's
core is slowly contracting and is  
finally expected to become a real core-collapse supernova (within a
few years) with a potentially very luminous display \citep{woo07}. 

\subsection{Was the historical SN 1961V similar to the 2012b eruptive event?} \label{sn61v}

The photometric comparison between SN 2009ip and SN 1961V shown in Figure \ref{fig2},  including the major eruption when SN 1961V reached an absolute peak magnitude of above -17, and the spectra \citep{bra71}
dominated by relatively narrow H lines, suggest a close similarity between these two transients,
hence supporting the statement that SN 1961V may have been another pulsational-pair instability event.

SN 1961V had a very troublesome genesis. For many years, from 1937 to 1954, its quiescent progenitor was the most luminous star in the host galaxy, NGC 1058. 
It had an apparent photographic magnitude of 18 \citep[][corresponding to an absolute mag M$_B \approx$ -12]{ber64}. 
With this luminosity, the star -likely an LBV- had an estimated M$_{ZAMS} >$ 80 M$_\odot$  \citep[adopting metallicities from 1/3 to 1 Z$_\odot$,][]{koc11}.
Then the object was observed at a constant magnitude of about 14.1-14.3 from July 1961 to November 1961,  and finally rose to a sharp maximum at mag $\approx$ 13 on December 11, 1961 \citep{ber64,ber67}.
 The peak was followed by a complex luminosity decline, which lasted for a few years with highly variable slopes (see Figure \ref{fig2}).

 The nature of this transient has been widely debated, and independent
 studies gave contradictory results on its real nature \citep[genuine
 SN or SN impostor; see discussion in][]{smi10}.  On the one hand,
 some authors state that a post-outburst surviving star \citep[known
 as ``Object 7'', see][and references therein]{van12} is visible in
 HST optical archival images.  The December 1961 light curve peak and
 the fluctuations in the post-maximum luminosity decline of SN 1961V
 could have been be produced by strong interaction between
 fast-moving, high-density material produced in an eruptive episode
 before 1961 with a lower-density, pre-existent circumstellar shell,
 without the need of a proper SN explosion, as also suggested by
 \citet{van12}. On the other hand, on the basis of the lack of
 sufficient infrared emission from the survived putative progenitor,
 \citet{koc11} proposed that SN 1961V had effectively exploded as a
 real SN, and its unusual observed properties could be explained via
 the ejecta interacting with a dense circumstellar medium.

 Even after half a century from the outburst, we can only speculate
 about the nature of SN 1961V, without giving definitive answers. 
Many
 years after that event, we tackle an analogous situation. 
SN 2009ip shares many strong similarities with SN 1961V, and
 the available information collected for SN 2009ip so far favor the
 pulsational pair-instability scenario of an extremely massive LBV.
 Whether the star has ended its life in the final core-collapse SN 
explosion, or the 2012b
 re-brightening event was due to shell-shell collisions is not known
 yet. Only long term monitoring of this erupting LBV will perhaps
 unveil its fate.

\section{Conclusions} \label{conc}

We presented the results of our spectroscopic and photometric observational
campaign for SN 2009ip, spanning a temporal window of more than 3 years.
There is clear evidence from the recent photometric history that, since August 2009, 
SN 2009ip was repeatedly seen in outburst. The light curve during the eruptive phases
was erratic and reached an absolute peak magnitude of -14 to -15. More recently,
another eruptive episode was observed (the 2012a event), lasting about 50 days
and reaching a luminosity of the same order of magnitude of previous outbursts 
(L $\approx$ 3 $\times$ 10$^{41}$ erg s$^{-1}$). However, since late September 2012,
another rebrightening was observed (the 2012b event), reaching an unprecedented peak luminosity
 (L $\approx$ 8 $\times$ 10$^{42}$ erg s$^{-1}$).

During all these outbursts, the spectra were dominated by strong and relatively narrow 
H emission lines, similar to those observed in several SN impostors, but also in type IIn
SNe. In addition broad P-Cygni absorptions indicative of high-velocity ejecta (up to 14-15000
km s$^{-1}$) became prominent during the 2012a event. We noticed, however, that the signature of high-velocity
material was observed in earlier spectra of SN 2009ip, suggesting that that presence
of fast ejecta does not necessarily imply a core-collapse SN. 

The recent spectro-photometric observations of SN 2009ip and its energetics favor a pulsational
pair-instability scenario where collisions among massive shells power the light curve, rather than a genuine SN explosion, 
although with the available information we cannot definitely 
rule out that the 2012a/b events witnessed the death of the LBV progenitor as a core-collapse SN.

Given this spectacular latest event, it would seem incumbent upon us
 to secure long-term monitoring campaigns (spectroscopy and imaging
 from both targeted and archival work) to track the variability
 history of the SN impostors. These long term campaigns are probably
 the most fruitful method to understanding the mechanisms that cause
 the unpredictable variability and determine the fate of LBVs, the
 most massive stars in the Local Universe. SN 2009ip should be one of 
best studied transient events in history.  Already, the data collected
on the progenitor star outstrips all the information we have on all
other SN progenitors to date. 

\acknowledgments

We thank the anonymous referee for a thorough reading of the manuscript and insightful comments. We are grateful to L. Girardi for useful discussions.
AP, EC, SB, MLP, AH, LT, and MT are partially supported by the PRIN-INAF 2011 with the project ”Transient Universe: from ESO Large to PESSTO”.
F.B. acknowledges support from FONDECYT through Postdoctoral grant 3120227.
M.H., G.P., F.B. acknowledge support by the Millennium Center for Supernova Science through
grant P10-064-F (funded by “Programa Bicentenario de Ciencia y Tecnolog\'ia
de CONICYT” and “Programa Iniciativa Cient\'ifica Milenio de MIDEPLAN”).
This work is supported in part under grant number 1108890 from the US National Science Foundation.

This work is partially based on observations of the European supernova collaboration involved in the ESO-NTT large programme 184.D-1140 led by Stefano Benetti.
It is also based on observations made with ESO VLT Telescopes at the Paranal Observatory under program IDs 087.D-0693 and 089.D-0325 (PI. S. Benetti), 083.D-0131 (PI. S. J. Smartt),
and with the ESO-NTT at the La Silla Observatory under program ID 083.D-0970 (PI. S. Benetti).

This paper is based on observations made with the Italian Telescopio Nazionale Galileo 
(TNG) operated on the island of La Palma by the Fundaci\'on Galileo Galilei of 
the INAF (Istituto Nazionale di Astrofisica). It is also based on observations made with the William Herschel Telescope (WHT) 
operated on the island of La Palma by the Isaac Newton Group in the Spanish Observatorio del Roque de los 
Muchachos of the Instituto de Astrofísica de Canarias; the Liverpool 
Telescope (LT) operated on the island of La Palma at the Spanish Observatorio del 
Roque de los Muchachos of the Instituto de Astrofisica de Canarias; the Nordic Optical Telescope (NOT), operated
on the island of La Palma jointly by Denmark, Finland, Iceland,
Norway, and Sweden, in the Spanish Observatorio del Roque de los
Muchachos of the Instituto de Astrof\'isica de Canarias;  the 2.2m telescope of  the Centro Astron\'omico Hispano Alem\'an 
(CAHA) at Calar Alto, operated jointly by the Max-Planck Institut f\"ur 
Astronomie and the Instituto de Astrof\'isica de Andaluc\'ia (CSIC); the 1.82m Copernico telescope of 
INAF-Asiago Observatory; the Gemini Observatory, which is operated by the Association of Universities for 
Research in Astronomy, Inc., under a cooperative agreement with the NSF on behalf of the Gemini partnership; and
the Panchromatic Robotic Optical Monitoring and Polarimetry (PROMPT) Telescopes, which are
 by the National Science Foundation, the University of North Carolina at Chapel Hill, Leonard Goodman, 
the National Aeronautics and Space Administration, Dudley Observatory, Henry Cox, and the Pisgah Astronomical Research Institute.
 The paper also presents observations obtained with the Faulkes Telescope South. The Faulkes Telescopes are maintained and
operated by Las Cumbres Observatory Global Telescope Network.

This publication makes use of data products from the Two Micron All Sky Survey, which is a joint project of the University of Massachusetts and the Infrared Processing and Analysis Center/California Institute of Technology, funded by the National Aeronautics and Space Administration and the National Science Foundation.

{\it Facilities:} \facility{NTT (ESO)}, \facility{VLT (ESO)}, \facility{TNG}, \facility{LT}, \facility{NOT}, \facility{Prompt}, \facility{Faulkes Telescope South},  \facility{Gemini}, \facility{SWIFT}.

\clearpage

\appendix

\section{Photometry Tables}  \label{appendix}

\begin{longtable}{cccccccc}
\tablewidth{0pt}
\tabletypesize{\footnotesize}
\tablecaption{Optical photometry of SN 2009ip. \label{tab_opt}}
\tablehead{ \colhead{Date}           & \colhead{JD-2400000}      &
\colhead{U}          & \colhead{B}  &
\colhead{V}          & \colhead{R}    &
\colhead{I}  & \colhead{Source}
}
30/08/09 &   55074.14 &         -        &           -         &          -         &           -        &    18.11  (0.13)   &   1  \\
30/08/09 &   55074.10 &         -        &           -         &          -         &    18.13  (0.09)   &           -        &   2  \\
30/08/09 &   55074.24 &         -        &           -         &    18.51  (0.18)   &           -        &           -        &   2 \\ 
06/09/09 &   55080.92 &         -        &           -         &    20.99  (0.37)   &           -        &           -        &   2  \\
07/09/09 &   55081.56 &         -        &           -         &    21.06  (0.05)   &    20.43  (0.03)   &           -        &   3  \\
07/09/09 &   55082.00 &         -        &           -         &    20.70  (0.17)   &           -        &           -        &   2  \\ 
07/09/09 &   55082.02 &         -        &           -         &    20.73  (0.20)   &           -        &           -        &   2  \\
07/09/09 &   55082.03 &         -        &           -         &          -         &    19.78  (0.24)   &           -        &   1  \\
08/09/09 &   55083.43 &         -        &           -         &          -         &    20.17  (0.08)   &           -        &   4  \\
09/09/09 &   55083.80 &         -        &           -         &    20.72  (0.32)   &           -        &           -        &   5  \\ 
09/09/09 &   55084.03 &         -        &           -         &    20.82  (0.25)   &           -        &           -        &   2   \\ 
09/09/09 &   55084.06 &         -        &           -         &          -         &    20.18  (0.20)   &           -        &   1  \\ 
10/09/09 &   55084.99 &         -        &           -         &    20.96  (0.27)   &           -        &           -        &   6  \\ 
10/09/09 &   55085.42 &         -        &           -         &          -         &    20.66  (0.07)   &           -        &   4  \\
12/09/09 &   55087.54 &         -        &           -         &    21.01  (0.10)   &    20.65  (0.09)   &    20.48  (0.19)   &   7  \\
13/09/09 &   55087.92 &         -        &           -         &          -         &           -        &   $>$19.61         &   1  \\ 
13/09/09 &   55088.10 &         -        &           -         &    21.01  (0.30)   &           -        &          -         &   6  \\ 
14/09/09 &   55088.81 &         -        &           -         &           -        &           -        &   $>$19.78         &   5  \\ 
14/09/09 &   55088.98 &         -        &           -         &           -        &           -        &    20.44  (0.30)   &   1  \\ 
14/09/09 &   55089.42 &         -        &           -         &           -        &    20.61  (0.06)   &           -        &   4  \\
14/09/09 &   55089.53 &         -        &           -         &    21.04  (0.03)   &    20.62  (0.05)   &    20.40  (0.19)   &   7  \\
23/09/09 &   55098.39 &         -        &           -         &           -        &    18.48  (0.06)   &           -        &   4  \\
24/09/09 &   55098.72 &         -        &           -         &    18.95  (0.13)   &           -        &           -        &   5  \\
24/09/09 &   55098.78 &         -        &           -         &           -        &           -        &    18.78  (0.21)   &   5  \\ 
24/09/09 &   55099.42 &         -        &   19.64  (0.03)     &    19.32  (0.04)   &    19.01  (0.04)   &    18.97  (0.10)   &   4  \\
25/09/09 &   55099.56 &         -        &           -         &    19.33  (0.02)   &    18.93  (0.06)   &           -        &   3  \\
25/09/09 &   55099.65 &         -        &           -         &    19.39  (0.12)   &           -        &           -        &   5  \\
26/09/09 &   55101.39 &         -        &   20.41  (0.11)     &    20.12  (0.13)   &    19.89  (0.07)   &    19.71  (0.16)   &   4  \\
27/09/09 &   55101.73 &         -        &           -         &    20.11  (0.25)   &           -        &           -        &   5  \\
29/09/09 &   55103.64 &         -        &           -         &    19.67  (0.03)   &    19.32  (0.03)   &           -        &   3  \\
02/10/09 &   55107.37 &         -        &           -         &           -        &    20.04  (0.14)   &           -        &   4  \\
06/10/09 &   55111.36 &         -        &           -         &           -        &    20.47  (0.13)   &           -        &   4  \\
08/10/09 &   55113.37 &         -        &           -         &           -        &    20.54  (0.05)   &           -        &   4  \\
10/10/09 &   55114.91 &         -        &           -         &    21.26  (0.36)   &           -        &           -        &   6  \\ 
13/10/09 &   55118.46 &         -        &           -         &    21.21  (0.10)   &    20.51  (0.06)   &    20.39  (0.10)   &   4  \\
18/10/09 &   55122.91 &         -        &           -         &    20.57  (0.31)   &           -        &           -        &   6  \\ 
22/10/09 &   55126.60 &         -        &    20.91 (0.03)     &    20.76  (0.09)   &    20.38  (0.17)   &    20.31  (0.16)   &   7  \\
30/10/09 &   55134.83 &         -        &           -         &            -       &    19.31  (0.04)   &           -        &   4  \\
02/11/09 &   55137.87 &         -        &           -         &            -       &    20.09  (0.06)   &           -        &   4  \\
06/11/09 &   55141.94 &         -        &           -         &    21.04  (0.28)   &           -        &           -        &   6  \\
07/11/09 &   55142.93 &         -        &           -         &    20.91  (0.18)   &           -        &           -        &   6  \\
10/11/09 &   55145.82 &         -        &           -         &            -       &    20.18  (0.03)   &           -        &  4  \\
13/11/09 &   55148.83 &         -        &           -         &            -       &    19.77  (0.02)   &           -        &  4  \\
21/11/09 &   55156.82 &         -        &           -         &            -       &    20.69  (0.06)   &           -        &  4  \\
22/11/09 &   55157.55 &         -        &           -         &            -       &    21.06  (0.04)   &           -        &  7  \\
24/11/09 &   55159.60 &         -        &    21.75 (0.06)     &    21.46  (0.10)   &    20.89  (0.09)   &    20.96  (0.16)   &  7  \\
29/11/09 &   55163.83 &         -        &           -         &            -       &    20.59  (0.09)   &           -        &  4  \\
05/12/09 &   55169.83 &         -        &           -         &            -       &    20.95  (0.07)   &           -        &  4  \\
04/10/10 &   55473.79 &         -        &           -         &            -       &    20.29  (0.15)   &           -        &  7  \\
06/10/10 &   55475.60 &         -        &           -         &            -       &    20.67  (0.07)   &           -        &  7  \\
06/10/10 &   55475.63 &         -        &           -         &            -       &    20.65  (0.03)   &           -        &  7  \\
06/10/10 &   55475.70 &  21.36 (0.06)    &    21.66 (0.02)     &     21.29 (0.08)   &    20.65  (0.08)   &    20.63  (0.12)   &  7  \\
28/10/10 &   55497.67 &         -        &           -         &            -       &    19.92  (0.03)   &           -        &  7  \\
30/10/10 &   55499.56 &         -        &    20.57 (0.02)     &     20.25 (0.05)   &    19.88  (0.06)   &           -        &  7  \\
30/10/10 &   55499.63 &         -        &           -         &            -       &           -        &    19.80  (0.12)   &  7  \\
20/12/10 &   55550.95 &         -        &           -         &            -       &    19.77  (0.10)   &           -        &  8 \\
23/12/10 &   55553.95 &         -        &           -         &            -       &    20.14  (0.16)   &           -        &  8 \\
31/12/10 &   55561.55 &         -        &           -         &            -       &    19.96  (0.12)   &           -        &  7  \\
02/01/11 &   55563.55 &         -        &           -         &            -       &    19.93  (0.08)   &           -        &  7  \\
25/03/11 &   55645.91 &         -        &           -         &            -       &    20.16  (0.40)   &           -        &  7  \\
11/04/11 &   55662.90 &         -        &           -         &            -       &    20.87  (0.20)   &           -        &  7  \\
10/05/11 &   55691.89 &         -        &           -         &            -       &    18.50  (0.07)   &           -        &  7  \\
25/05/11 &   55707.13 &         -        &           -         &            -       &    20.59  (0.15)   &           -        &  8 \\
26/05/11 &   55708.20 &         -        &           -         &            -       &    20.83  (0.11)   &           -        &  8 \\
27/05/11 &   55709.19 &         -        &           -         &            -       &    20.81  (0.18)   &           -        &  8 \\
04/06/11 &   55717.27 &         -        &           -         &            -       &    18.18  (0.07)   &           -        &  8 \\
10/06/11 &   55723.26 &         -        &           -         &            -       &    20.12  (0.17)   &           -        &  8 \\
19/06/11 &   55732.22 &         -        &           -         &            -       &    20.48  (0.22)   &           -        &  8 \\
24/06/11 &   55736.82 &         -        &           -         &            -       &    20.63  (0.10)   &           -        &  7  \\
24/06/11 &   55737.17 &         -        &           -         &            -       &    20.60  (0.07)   &           -        &  8 \\
26/06/11 &   55738.78 &         -        &           -         &            -       &    20.87  (0.09)   &           -        &  7  \\
03/07/11 &   55746.06 &         -        &           -         &            -       &    20.48  (0.16)   &           -        &  8 \\
08/07/11 &   55751.11 &         -        &           -         &            -       &    18.18  (0.05)   &           -        &  8 \\
14/07/11 &   55757.02 &         -        &           -         &            -       &    20.03  (0.41)   &           -        &  8 \\
23/07/11 &   55766.20 &         -        &           -         &            -       &    20.34  (0.09)   &           -        &  8 \\
28/07/11 &   55771.21 &         -        &           -         &            -       &    20.45  (0.08)   &           -        &  8 \\
03/08/11 &   55777.24 &         -        &           -         &            -       &    18.46  (0.05)   &           -        &  8 \\
24/08/11 &   55797.93 &         -        &           -         &            -       &    19.40  (0.05)   &           -        &  8 \\
27/08/11 &   55801.12 &         -        &           -         &            -       &    19.91  (0.03)   &           -        &  8 \\
28/08/11 &   55802.13 &         -        &           -         &            -       &    20.17  (0.05)   &           -        &  8 \\
30/08/11 &   55804.24 &         -        &           -         &            -       &    20.39  (0.05)   &           -        &  8 \\
01/09/11 &   55805.90 &         -        &           -         &            -       &    20.49  (0.14)   &           -        &  8 \\
03/09/11 &   55808.01 &         -        &           -         &            -       &    20.73  (0.07)   &           -        &  8 \\
05/09/11 &   55809.90 &         -        &           -         &            -       &    20.89  (0.15)   &           -        &  8 \\
06/06/11 &   55810.66 &         -        &           -         &            -       &    21.00  (0.08)   &           -        &  7  \\
07/09/11 &   55811.04 &         -        &           -         &            -       &    21.14  (0.25)   &           -        &  8 \\
10/09/11 &   55815.02 &         -        &           -         &            -       &    $>$19.67        &           -        &  8  \\
13/09/11 &   55817.90 &         -        &           -         &            -       &    20.77  (0.17)   &           -        &  8 \\
16/09/11 &   55821.18 &         -        &           -         &            -       &    20.29  (0.11)   &           -        &  8 \\
17/09/11 &   55821.92 &         -        &           -         &            -       &    19.56  (0.05)   &           -        &  8 \\
19/09/11 &   55823.94 &         -        &           -         &            -       &    18.57  (0.03)   &           -        &  8 \\
21/09/11 &   55826.15 &         -        &           -         &            -       &    17.86  (0.04)   &           -        &  8 \\
22/09/11 &   55826.91 &         -        &           -         &            -       &    18.48  (0.04)   &           -        &  8 \\
22/09/11 &   55826.98 &         -        &           -         &            -       &    18.58  (0.03)   &           -        &  8 \\
22/09/11 &   55827.09 &         -        &           -         &            -       &    18.72  (0.03)   &           -        &  8 \\
24/09/11 &   55829.18 &         -        &           -         &            -       &    19.02  (0.24)   &           -        &  8 \\
26/09/11 &   55830.90 &         -        &           -         &            -       &    19.36  (0.07)   &           -        &  8 \\
26/09/11 &   55830.98 &         -        &           -         &            -       &    19.40  (0.03)   &           -        &  8 \\
26/09/11 &   55831.13 &         -        &           -         &            -       &    19.45  (0.06)   &           -        &  8 \\
27/09/11 &   55831.99 &         -        &           -         &            -       &    19.65  (0.07)   &           -        &  8 \\
30/09/11 &   55834.90 &         -        &           -         &            -       &    20.02  (0.05)   &           -        &  8 \\
04/10/11 &   55838.91 &         -        &           -         &            -       &    18.85  (0.04)   &           -        &  8 \\
09/10/11 &   55843.90 &         -        &           -         &            -       &  $>$18.45          &           -        &  8  \\
18/10/11 &   55852.52 &         -        &           -         &            -       &    20.87  (0.09)   &           -        &  7  \\
18/10/11 &   55852.99 &         -        &           -         &            -       &    20.82  (0.08)   &           -        &  8 \\
20/10/11 &   55854.59 &         -        &           -         &            -       &    20.90  (0.14)   &           -        &  7  \\
21/10/11 &   55855.91 &         -        &           -         &            -       &    20.84  (0.07)   &           -        &  8 \\
22/10/11 &   55856.02 &         -        &           -         &            -       &    20.84  (0.08)   &           -        &  8 \\
24/10/11 &   55858.91 &         -        &           -         &            -       &    20.50  (0.06)   &           -        &  8 \\
17/11/11 &   55882.57 &         -        &           -         &            -       &    20.77  (0.10)   &           -        &  7  \\
27/11/11 &   55892.94 &         -        &           -         &            -       &    19.96  (0.06)   &           -        &  8  \\
19/12/11 &   55914.54 &         -        &           -         &            -       &    19.85  (0.09)   &           -        &  7  \\
21/12/11 &   55916.53 &         -        &           -         &            -       &    19.70  (0.11)   &           -        &  7  \\
23/04/12 &   56040.43 &         -        &           -         &            -       &    19.92  (0.13)   &           -        &  7  \\
08/08/12 &   56147.73 &  17.54 (0.01)    &   18.51   (0.01)    &     18.43 (0.04)   &    18.22  (0.03)   &    18.12  (0.05)   &  7  \\
09/08/12 &   56148.91 &         -        &           -         &            -       &    18.46  (0.08)   &           -        &  7  \\
11/08/12 &   56150.90 &         -        &           -         &            -       &    18.23  (0.05)   &           -        &  7  \\
18/08/12 &   56157.76 &  16.54 (0.22)    &   17.24   (0.10)    &     17.05 (0.11)   &    16.86  (0.06)   &    16.73  (0.11)   &  9  \\
25/08/12 &   56164.78 &         -        &           -         &            -       &    16.94  (0.02)   &           -        &  7  \\
26/08/12 &   56165.59 &  16.09 (0.01)    &   16.88   (0.01)    &     16.84 (0.02)   &    16.61  (0.02)   &    16.53  (0.02)   &  7  \\
27/08/12 &   56166.65 &  16.24 (0.01)    &   17.02   (0.01)    &     16.93 (0.01)   &    16.73  (0.03)   &    16.58  (0.03)   &  7  \\
29/08/12 &   56168.54 &         -        &           -         &            -       &    16.56  (0.05)   &           -        &  10  \\
31/08/12 &   56170.54 &  16.32 (0.02)    &   17.02   (0.01)    &     16.84 (0.02)   &    16.54  (0.05)   &    16.48  (0.08)   &  10  \\
01/09/12 &   56171.60 &         -        &   16.92   (0.03)    &     16.77 (0.03)   &    16.52  (0.04)   &    16.49  (0.05)   &  11  \\
02/09/12 &   56172.62 &         -        &   17.02   (0.08)    &     16.73 (0.02)   &    16.57  (0.04)   &    16.41  (0.04)   &  11  \\
05/09/12 &   56175.57 &  16.72 (0.03)    &   17.31   (0.06)    &     17.09 (0.04)   &    16.82  (0.03)   &    16.77  (0.03)   &  12  \\
05/09/12 &   56175.85 &         -        &           -         &     16.99 (0.07)   &    16.78  (0.07)   &    16.79  (0.10)   &  11  \\
06/09/12 &   56176.56 &         -        &   17.34   (0.06)    &     17.08 (0.03)   &    16.85  (0.04)   &    16.70  (0.05)   &  11  \\
07/09/12 &   56177.53 &         -        &           -         &            -       &    16.80  (0.08)   &           -        &  7  \\
07/09/12 &   56177.60 &  16.39 (0.12)    &           -         &            -       &           -        &           -        &  7  \\
10/09/12 &   56180.53 &         -        &           -         &            -       &    16.79  (0.10)   &           -        &  7  \\
11/09/12 &   56181.67 &         -        &          -          &     16.98 (0.05)   &    16.58  (0.04)   &    16.57  (0.07)   &  13 \\
18/09/12 &   56188.54 &         -        &           -         &            -       &    17.66  (0.10)   &           -        &  7  \\
22/09/12 &   56193.46 &  17.96 (0.02)    &   18.23   (0.04)    &     18.00 (0.04)   &    17.58  (0.09)   &    17.58  (0.06)   &  10  \\
23/09/12 &   56193.59 &         -        &   18.22   (0.07)    &     17.97 (0.04)   &    17.61  (0.06)   &    17.48  (0.07)   &  11  \\
23/09/12 &   56194.10 &         -        &           -         &            -       &           -        &    17.32  (0.10)   &   2 \\
24/09/12 &   56194.58 &         -        &   18.14   (0.06)    &     17.84 (0.03)   &    17.53  (0.06)   &    17.45  (0.06)   &  11  \\
24/09/12 &   56194.66 &         -        &          -          &     17.77 (0.08)   &    17.12  (0.06)   &           -        &  13 \\
24/09/12 &   56195.09 &         -        &           -         &            -       &    16.32  (0.05)   &    16.53  (0.07)   &  2,6  \\
25/09/12 &   56195.63 &         -        &           -         &     15.62 (0.02)   &    15.48  (0.05)   &    15.43  (0.04)   &  11  \\
25/09/12 &   56195.70 &         -        &           -         &            -       &           -        &    15.03  (0.12)   &   5  \\
25/09/12 &   56196.01 &         -        &           -         &            -       &    15.00  (0.03)   &    15.04  (0.06)   &  2,6  \\
26/09/12 &   56196.57 &         -        &           -         &     14.91 (0.06)   &           -        &           -        &  14  \\    
26/09/12 &   56196.60 &         -        &           -         &     14.84 (0.02)   &    14.65  (0.03)   &    14.67  (0.04)   &  11  \\
26/09/12 &   56196.98 &         -        &           -         &            -       &    14.52  (0.05)   &    14.56  (0.05)   &  2,6  \\
27/09/12 &   56197.57 &         -        &           -         &     14.59 (0.08)   &           -        &    14.45  (0.13)   &  14  \\    
27/09/12 &   56197.63 &         -        &           -         &     14.52 (0.02)   &    14.37  (0.03)   &    14.39  (0.04)   &  11  \\
27/09/12 &   56198.03 &         -        &           -         &            -       &           -        &    14.33  (0.08)   &  2  \\
28/09/12 &   56198.56 &         -        &           -         &     14.39 (0.06)   &           -        &    14.19  (0.07)   &  14  \\    
28/09/12 &   56198.80 &         -        &   14.32   (0.03)    &     14.32 (0.02)   &    14.09  (0.04)   &    14.12  (0.03)   &  11  \\
28/09/12 &   56198.93 &         -        &           -         &            -       &    14.14  (0.05)   &    14.21  (0.07)   &  2,6  \\
28/09/12 &   56199.42 &  13.20 (0.03)    &   14.20   (0.02)    &     14.28 (0.02)   &    14.11  (0.01)   &    14.11  (0.02)   &  4  \\
29/09/12 &   56199.65 &         -        &   14.28   (0.02)    &     14.20 (0.01)   &    14.09  (0.02)   &    14.08  (0.02)   &  11  \\
29/09/12 &   56199.70 &         -        &           -         &     14.23 (0.06)   &           -        &    13.96  (0.07)   &  14  \\    
29/09/12 &   56200.08 &         -        &           -         &            -       &    14.05  (0.02)   &           -        &  15  \\
29/09/12 &   56200.43 &  13.03 (0.03)    &   14.09   (0.02)    &     14.15 (0.02)   &    13.98  (0.01)   &    13.98  (0.02)   &  4  \\
30/09/12 &   56200.57 &         -        &           -         &     14.16 (0.07)   &           -        &    13.99  (0.06)   &  14  \\    
30/09/12 &   56201.11 &         -        &           -         &            -       &    14.08  (0.02)   &           -        &  15  \\
30/09/12 &   56201.19 &         -        &           -         &            -       &    13.91  (0.04)   &    13.98  (0.03)   &  2,6  \\
01/10/12 &   56201.53 &         -        &   14.02   (0.04)    &     14.03 (0.02)   &    13.91  (0.02)   &    13.95  (0.01)   &  11  \\
01/10/12 &   56201.56 &         -        &           -         &     14.02 (0.05)   &           -        &    13.91  (0.04)   &  14  \\    
01/10/12 &   56201.94 &         -        &           -         &            -       &    13.85  (0.03)   &    13.94  (0.03)   &  2,6  \\
01/10/12 &   56202.04 &         -        &           -         &            -       &    13.91  (0.02)   &           -        &  15  \\
01/10/12 &   56202.44 &  12.98 (0.03)    &   13.97   (0.02)    &     14.03 (0.02)   &    13.86  (0.01)   &    13.88  (0.02)   &  4  \\ 
02/10/12 &   56202.55 &         -        &           -         &     13.99 (0.07)   &           -        &    13.81  (0.08)   &  14  \\    
02/10/12 &   56202.76 &         -        &           -         &     13.96 (0.05)   &           -        &           -        &  5  \\  
03/10/12 &   56203.55 &         -        &           -         &     13.96 (0.06)   &           -        &    13.88  (0.08)   &  14  \\    
03/10/12 &   56203.72 &         -        &           -         &     13.86 (0.03)   &           -        &           -        &  5  \\  
04/10/12 &   56204.52 &         -        &   13.95   (0.05)    &     13.75 (0.02)   &    13.78  (0.03)   &    13.79  (0.05)   &  11  \\ 
04/10/12 &   56204.55 &         -        &           -         &     13.90 (0.05)   &           -        &    13.80  (0.07)   &  14  \\    
04/10/12 &   56205.11 &         -        &           -         &            -       &    13.74  (0.01)   &           -        &  15  \\
05/10/12 &   56205.55 &         -        &           -         &     13.81 (0.05)   &           -        &    13.72  (0.07)   &  14  \\    
05/10/12 &   56205.61 &         -        &           -         &     13.73 (0.05)   &           -        &           -        &  5  \\  
05/10/12 &   56206.09 &         -        &           -         &            -       &    13.70  (0.01)   &           -        &  15  \\
06/10/12 &   56206.54 &         -        &           -         &     13.80 (0.05)   &           -        &    13.68  (0.05)   &  14  \\    
06/10/12 &   56206.71 &         -        &           -         &     13.71 (0.03)   &           -        &           -        &  5  \\  
06/10/12 &   56207.09 &         -        &           -         &            -       &    13.70  (0.01)   &           -        &  15  \\
06/10/12 &   56207.40 &  12.82 (0.04)    &   13.80   (0.02)    &     13.82 (0.02)   &    13.65  (0.02)   &    13.66  (0.02)   &  4  \\
07/10/12 &   56208.41 &         -        &          -          &     13.83 (0.02)   &    13.65  (0.01)   &    13.67  (0.02)   &  4  \\
08/10/12 &   56208.54 &         -        &           -         &     13.83 (0.05)   &           -        &    13.69  (0.07)   &  14  \\    
08/10/12 &   56209.42 &         -        &          -          &     13.88 (0.02)   &    13.73  (0.01)   &    13.68  (0.02)   &  4  \\
09/10/12 &   56209.60 &         -        &          -          &     13.88 (0.03)   &    13.75  (0.04)   &    13.69  (0.07)   &  13 \\
09/10/12 &   56209.70 &         -        &           -         &     13.86 (0.05)   &           -        &    13.63  (0.09)   &  14  \\    
09/10/12 &   56210.07 &         -        &           -         &            -       &    13.74  (0.01)   &           -        &  15  \\
09/10/12 &   56210.43 &  12.99 (0.03)    &   13.91   (0.02)    &     13.92 (0.01)   &    13.73  (0.01)   &    13.70  (0.02)   &  4  \\
10/10/12 &   56210.53 &         -        &           -         &     13.87 (0.06)   &           -        &    13.71  (0.07)   &  14  \\    
10/10/12 &   56210.65 &         -        &   13.95   (0.03)    &            -       &           -        &           -        &  11  \\
10/10/12 &   56211.08 &         -        &           -         &            -       &    13.76  (0.01)   &           -        &  15  \\
11/10/12 &   56211.65 &         -        &           -         &     13.91 (0.03)   &           -        &    13.61  (0.07)   &  14  \\    
11/10/12 &   56212.39 &  13.04 (0.03)    &   13.96   (0.02)    &     13.90 (0.02)   &    13.73  (0.01)   &    13.69  (0.02)   &  4  \\
12/10/12 &   56212.53 &         -        &           -         &     13.87 (0.06)   &           -        &    13.67  (0.06)   &  14  \\    
13/10/12 &   56213.53 &         -        &           -         &     13.87 (0.01)   &    13.74  (0.03)   &    13.66  (0.05)   &  11 \\
13/10/12 &   56213.58 &         -        &           -         &     13.86 (0.06)   &           -        &    13.59  (0.11)   &  14  \\    
13/10/12 &   56214.38 &  13.02 (0.03)    &   13.96   (0.02)    &     13.91 (0.02)   &    13.71  (0.01)   &    13.68  (0.02)   &  4  \\
15/10/12 &   56216.11 &         -        &           -         &            -       &    13.79  (0.01)   &           -        &  15  \\
16/10/12 &   56216.73 &         -        &   14.11   (0.02)    &     14.00 (0.03)   &    13.79  (0.02)   &    13.74  (0.01)   &  11  \\
\tablenotetext{1}{0.3-m Mewlon Telescope + ST10 XME camera, Coral Tower Observatory, Cairns (Australia)}
\tablenotetext{2}{0.41-m RCOS Telescope + STL6K camera, Coral Tower Observatory, Cairns (Australia)}
\tablenotetext{3}{8.2-m Very Large Telescope UT1 + FORS2, European Southern Observatory - Cerro Paranal (Chile)}
\tablenotetext{4}{2-m Liverpool Telescope + RATCam, La Palma, Canary Islands (Spain)}
\tablenotetext{5}{0.51-m RCOS Telescope + STL11K camera, New Mexico Skies, Mayhill, New Mexico (USA)}
\tablenotetext{6}{0.33-m RCOS Telescope + STL6K camera, in 2009 at the  Macedon Ranges Observatory, Melbourne; in 2012 at the Coral Tower Observatory, Cairns (Australia)} 
\tablenotetext{7}{3.58-m New Technology Telescope + EFOSC2, European Southern Observatory - La Silla (Chile)}
\tablenotetext{8}{2-m Faulkes Telescope South + EM03, Siding Spring Observatory (Australia)}
\tablenotetext{9}{8.2-m Very Large Telescope UT2 + XShooter (spectro-photometry), European Southern Observatory - Cerro Paranal (Chile)}
\tablenotetext{10}{3.58-m Telescopio Nazionale Galileo + Dolores, La Palma, Canary Islands (Spain)}
\tablenotetext{11}{0.41-m Panchromatic Robotic Optical Monitoring and Polarimetry Telescopes (PROMPT) (3+5), at Cerro Tololo Inter-American Observatory (Chile)}
\tablenotetext{12}{2.56-m Nordic Optical Telescope + ALFOSC, La Palma, Canary Islands (Spain)}
\tablenotetext{13}{0.51-m Cassegrain Reflector + Apogee U42 CCD Camera, Barber Observatory, Pleasant Plains, Illinois (USA)}
\tablenotetext{14}{0.40-m Optimized Dall-Kirkham + FLI Kodak 16803 CCD, Remote Observatory Atacama Desert, San Pedro de Atacama (Chile)}
\tablenotetext{15}{0.235-m Perth Exoplanet Survey Telescope + SBIG ST-8XME CCD, Perth (Australia)}

\end{longtable}

\clearpage

\begin{deluxetable}{cccccccc}
\tabletypesize{\footnotesize}
\tablecaption{Swift/UVOT photometry of SN 2009ip. \label{tab_swift}}
\tablehead{ \colhead{Date}           & \colhead{JD}      &
\colhead{UVw2}          & \colhead{UVm2}  &
\colhead{UVw1}          & \colhead{u}    &
\colhead{b}   & \colhead{v}  }
\startdata
10/09/09 &  55085.17 &   $>$21.64      &  $>$21.10     &  $>$21.18      &    $>$20.87     &  $>$21.22     &  $>$20.30     \\
04/09/12 &  56175.40 &   18.00 (0.08)  &        -      &  17.20 (0.05)  &    16.49 (0.04) &  17.29 (0.04) &  17.01 (0.02) \\
06/09/12 &  56177.14 &   18.02 (0.08)  &  17.95 (0.08) &  17.06 (0.07)  &    16.51 (0.06) &  17.33 (0.07) &  17.23 (0.12) \\
13/09/12 &  56183.94 &   19.38 (0.32)  &  19.04 (0.16) &  18.29 (0.12)  &    17.31 (0.07) &  17.73 (0.08) &  17.83 (0.14) \\ 
20/09/12 &  56191.25 &         -       &        -      &  19.30 (0.24)  &    17.91 (0.13) &  18.50 (0.17) &        -      \\
22/09/12 &  56193.26 &   20.58 (0.26)  &  20.52 (0.32) &  19.36 (0.23)  &    17.91 (0.07) &  18.24 (0.06) &  18.17 (0.07) \\
26/09/12$\ast$ &  56196.73 &   12.85 (0.07)  &  12.82 (0.06) &  13.00 (0.05)  &    13.43 (0.04) &  14.78 (0.04) &  14.90 (0.04) \\  
27/09/12 &  56197.95 &   12.61 (0.04)  &       -       &  12.78 (0.04)  &         -       &       -       &        -       \\
27/09/12 &  56198.20 &   12.50 (0.04)  &       -       &  12.68 (0.04)  &    13.02 (0.04) &  14.40 (0.04) &        -       \\
28/09/12 &  56198.54 &   12.47 (0.04)  &  12.36 (0.04) &  12.66 (0.04)  &    12.90 (0.04) &  14.40 (0.05) &  14.58 (0.06)  \\
28/09/12 &  56199.09 &   12.39 (0.04)  &       -       &  12.63 (0.04)  &         -       &       -       &        -       \\
29/09/12 &  56199.96 &        -       &       -        &  12.57 (0.04)  &         -       &       -       &        -       \\   
29/09/12 &  56200.15 &   12.28 (0.04)  &  12.38 (0.05) &  12.45 (0.04)  &    12.81 (0.04) &  14.11 (0.04) &  14.22 (0.04)  \\
30/09/12 &  56200.74 &        -        &       -       &       -        &    12.75 (0.04) &       -       &        -       \\
30/09/12 &  56201.01 &   12.21 (0.04)  &       -       &       -        &         -       &       -       &        -       \\
30/09/12 &  56201.49 &   12.18 (0.04)  &  12.07 (0.04) &  12.35 (0.04)  &    12.67 (0.04) &  13.94 (0.04) &  14.12 (0.04)  \\
01/10/12 &  56201.55 &   12.19 (0.04)  &       -       &       -        &         -       &       -       &        -       \\
01/10/12 &  56201.93 &   12.20 (0.04)  &       -       &       -        &         -       &       -       &        -       \\
01/10/12 &  56202.49 &   12.16 (0.04)  &  12.10 (0.04) &  12.36 (0.04)  &    12.65 (0.04) &  13.99 (0.04) &  14.04 (0.04)  \\
02/10/12 &  56202.76 &   12.18 (0.04)  &  12.11 (0.04) &  12.36 (0.04)  &    12.65 (0.04) &  13.95 (0.04) &  14.08 (0.04)  \\
03/10/12 &  56203.56 &   12.20 (0.04)  &  12.11 (0.04) &  12.36 (0.05)  &    12.65 (0.05) &  13.96 (0.04) &  14.09 (0.04)  \\   
03/10/12 &  56203.79 &   12.23 (0.04)  &       -       &       -        &         -       &       -       &       -        \\   
04/10/12 &  56204.55 &   12.14 (0.04)  &  12.04 (0.04) &  12.30 (0.05)  &    12.63 (0.05) &  13.86 (0.04) &  13.92 (0.04)  \\   
04/10/12 &  56204.89 &   12.10 (0.04)  &       -       &      -         &        -        &       -       &       -        \\ 
04/10/12 &  56205.01 &        -        &       -       &      -         &        -        &  13.80 (0.12) &       -        \\ 
05/10/12 &  56205.90 &   12.10 (0.04)  &       -       &      -         &        -        &       -       &       -        \\ 
05/10/12 &  56206.02 &        -        &       -       &      -         &        -        &  13.77 (0.10) &       -        \\ 
06/10/12 &  56206.62 &   12.10 (0.04)  &  12.02 (0.04) &  12.24 (0.04)  &   12.52 (0.05)  &  13.75 (0.05) &  13.86 (0.04)  \\ 
06/10/12 &  56206.70 &   12.13 (0.04)  &       -       &      -         &        -        &       -       &       -        \\ 
07/10/12 &  56207.93 &   12.27 (0.04)  &       -       &      -         &        -        &       -       &       -        \\ 
08/10/12 &  56209.17 &   12.42 (0.04)  &  12.26 (0.04) &  12.48 (0.05)  &   12.62 (0.05)  &  13.84 (0.05) &  13.87 (0.04)  \\
10/10/12 &  56210.56 &   12.61 (0.05)  &  12.44 (0.04) &  12.59 (0.05)  &   12.80 (0.05)  &  13.92 (0.05) &  13.88 (0.04)  \\
12/10/12 &  56212.70 &   12.76 (0.05)  &  12.55 (0.05) &  12.74 (0.05)  &   12.77 (0.05)  &  13.91 (0.05) &  13.94 (0.05)  \\
14/10/12 &  56214.63 &   12.90 (0.05)  &  12.68 (0.05) &  12.82 (0.05)  &   12.86 (0.05)  &  13.98 (0.05) &  13.93 (0.05)  \\
16/10/12 &  56216.71 &   13.39 (0.05)  &  13.12 (0.05) &  13.33 (0.05)  &   13.07 (0.05)  &  14.14 (0.05) &  14.07 (0.05)  \\
\enddata
\tablecomments{The photometry at the epoch marked with $\ast$ has been published by \protect\citet{marg12b}.}
\end{deluxetable}

\begin{deluxetable}{cccccc}
\tabletypesize{\footnotesize}
\tablecaption{Near-infrared photometry of SN 2009ip obtained during the 2012a and 2012b events. \label{tab_NIR}}
\tablehead{ \colhead{Date}           & \colhead{JD}      &
\colhead{J}          & \colhead{H}  &
\colhead{Ks}        & \colhead{Source}
}
\startdata
18/08/12 &  56157.76  & 16.32  (0.177)     & 16.16 (0.16)        & 16.60 (0.22)   & 1    \\
26/08/12 &  56165.80  & 16.25  (0.098)     & 16.11 (0.12)        & 15.88 (0.15)   & 2    \\
08/09/12 &  56178.53  & 16.47  (0.132)     & 16.28 (0.13)        & 15.92 (0.18)   & 3    \\
09/09/12 &  56179.60  & 16.46  (0.102)     & 16.28 (0.12)        & 16.02 (0.15)   & 2    \\
22/09/12 &  56193.59  & 17.17  (0.120)     & 17.15 (0.18)        & 16.86 (0.17)   & 2    \\ 
27/09/12 &  56198.48  & 14.17  (0.097)     & 13.92 (0.11)        & 13.48 (0.16)   & 3    \\
28/09/12 &  56199.47  & 14.06  (0.108)     & 13.78 (0.13)        & 13.38 (0.17)   & 3    \\
01/10/12 &  56202.46  & 13.97  (0.110)     & 13.73 (0.14)        & 13.29 (0.18)   & 3    \\ 
12/10/12 &  56212.75  & 13.65  (0.184)     & 13.35 (0.17)        & 13.38 (0.27)   & 4    \\
\enddata
\tablenotetext{1}{8.2-m Very Large Telescope UT2 + XShooter (spectro-photometry), European Southern Observatory - Cerro Paranal (Chile)}
\tablenotetext{2}{3.58-m New Technology Telescope + SOFI, European Southern Observatory - La Silla (Chile)}
\tablenotetext{3}{2.56-m Nordic Optical Telescope + NOTCam, La Palma, Canary Islands (Spain)}
\tablenotetext{4}{0.6-m Rapid Eye Mount + REMIR, La Silla (Chile)}
\end{deluxetable}

\clearpage

\begin{deluxetable}{cccccccc}
\tabletypesize{\footnotesize}
\tablecaption{Unpublished optical photometry of NGC3432-LBV1 (aka SN 2000ch). \label{tab_00ch}}
\tablehead{ \colhead{Date}           & \colhead{JD-2400000}      &
\colhead{U}          & \colhead{B}  &
\colhead{V}          & \colhead{R}    &
\colhead{I}  & \colhead{Source} 
}
\startdata
02/07/10  &  55380.37  &       -        &       -        &       -        &  19.78 (0.05)  &       -         &    1   \\
14/07/10  &  55392.37  &       -        &       -        &       -        &  19.09 (0.05)  &       -         &    1   \\ 
20/07/10  &  55398.36  &       -        &       -        &       -        &  20.53 (0.31)  &       -         &    1   \\ 
10/10/10  &  55489.70  &       -        &       -        &       -        &  19.74 (0.10)  &       -         &    1   \\ 
26/10/10  &  55495.66  &       -        & 22.25  (0.40)  & 20.97  (0.17)  &  19.99 (0.06)  & 20.17  (0.22)   &    1   \\ 
17/11/10  &  55517.71  &       -        &       -        &       -        &  19.93 (0.11)  &       -         &    1   \\ 
20/11/10  &  55520.65  &       -        &       -        &       -        &  20.38 (0.09)  &       -         &    1   \\ 
15/01/11  &  55576.75  &       -        &       -        &       -        &  19.14 (0.05)  &       -         &    1   \\ 
08/02/11  &  55600.70  &       -        &       -        &       -        &  19.98 (0.07)  &       -         &    2    \\
09/02/11  &  55601.53  &       -        & 21.21  (0.12)  & 20.90  (0.10)  &  20.05 (0.09)  & 19.97  (0.15)   &    3   \\	
09/02/11  &  55602.49  &       -        &       -        &       -        &  20.09 (0.05)  &       -         &    3   \\ 
22/02/11  &  55615.77  &       -        &       -        &       -        &  19.97 (0.05)  &       -         &    4    \\
06/04/11  &  55658.49  &       -        & 21.26  (0.15)  & 21.25  (0.19)  &  20.28 (0.18)  & $>$19.99        &    3   \\
10/04/11  &  55662.40  &       -        &       -        &       -        &  20.23 (0.12)  &       -         &    1   \\ 
01/05/11  &  55683.40  &       -        &       -        &       -        &  $>$19.54      &       -         &    3   \\    
07/05/11  &  55689.48  &       -        &       -        &       -        &  19.42 (0.04)  &       -         &    1   \\ 
22/05/11  &  55704.35  &       -        &       -        &       -        &  19.17 (0.05)  &       -         &    1   \\
30/06/11  &  55743.37  &       -        &       -        &       -        &  19.96 (0.11)  &       -         &    1   \\ 
01/11/11  &  55866.56  &    $>$19.61    &    $>$22.21    &    $>$21.35    &  20.48 (0.23)  &     $>$20.22    &    3   \\  
02/11/11  &  55867.69  &       -        &       -        & 21.80  (0.36)  &  20.53 (0.19)  & 20.91  (0.28)   &    1   \\
16/11/11  &  55881.72  &       -        &       -        &       -        &  20.95 (0.33)  &       -         &    1   \\ 
20/11/11  &  55885.63  &       -        &       -        &       -        &  21.18 (0.20)  &       -         &    3   \\
22/12/11  &  55917.62  &       -        &       -        &       -        &  21.07 (0.07)  &       -         &    2   \\
24/12/11  &  55919.57  &       -        &       -        &       -        &  21.42 (0.19)  &       -         &    1   \\
21/01/12  &  55948.41  &       -        &       -        &       -        &  20.12 (0.47)  &       -         &    3   \\
30/01/12  &  55957.47  &       -        &       -        &       -        &  19.87 (0.22)  &       -         &    3   \\
17/03/12  &  56003.51  &       -        &       -        &       -        &  19.40 (0.07)  &       -         &    3   \\
26/03/12  &  56013.36  &       -        &       -        &       -        &  19.13 (0.08)  &       -         &    3   \\
28/03/12  &  56015.45  &       -        &       -        &       -        &  19.23 (0.08)  &       -         &    3   \\
17/07/12  &  56126.34  &       -        &       -        &       -        &  $>$20.90      &       -         &    3   \\
26/07/12  &  56135.35  &       -        &       -        &       -        &  $>$20.70      &       -         &    3   \\ 
26/10/12  &  56226.26  &       -        &    $>$20.12    &    $>$20.16    &  $>$19.82      &    $>$19.43     &    5   \\ 
06/11/12  &  56237.63  &       -        &       -        &       -        &  $>$21.35      &       -         &    3   \\ 
08/11/12  &  56239.66  &       -        &       -        &       -        &  21.80 (0.24)  &       -         &    3   \\ 
05/12/12  &  56266.73  &       -        &       -        &       -        &  $>$20.97      &       -         &    3   \\ 
19/12/12  &  56280.59  &       -        &  20.33 (0.10)  &   19.96 (0.09) &  19.45 (0.07)  &    19.05 (0.11) &    3   \\
\enddata
\tablenotetext{1}{2.2-m Calar Alto Telescope + CAFOS, Calar Alto, Almeria (Spain)}
\tablenotetext{2}{2.56-m Nordic Optical Telescope + ALFOSC, La Palma, Canary Islands (Spain)}
\tablenotetext{3}{1.82-m Copernico Telescope + AFOSC, Mt. Ekar, Asiago (Italy)} 
\tablenotetext{4}{4.2-m William Herschel Telescope + ACAM, La Palma, Canary Islands (Spain)} 
\tablenotetext{5}{2.0-m Lverpool Telescope + RATCam, La Palma, Canary Islands (Spain)}  
\end{deluxetable}

\clearpage

\begin{sidewaystable}
\caption{Optical, SWIFT/UVOT and NIR magnitudes of the reference stars in the field of SN 2009ip. The NIR magnitudes are take from the 2-MASS catalogue. }
\centering
\begin{tabular}{cccccccccccc} \hline\hline
Filter   & Star 1   & Star 2       & Star 3       & Star 4       &  Star 5      & Star 6       & Star 7       & Star 8       & Star 9       & Star 10      &Star 11 \\ \hline
 U   & 15.70 (0.01) & 16.14 (0.04) &   -          &    -         & 21.06 (0.07) & 20.44 (0.05) &     -        &     -        &    -         & 20.33 (0.02) & 19.37 (0.02) \\
 B   & 15.76 (0.02) & 16.20 (0.01) & 21.13 (0.02) & 19.10 (0.05) & 20.14 (0.04) & 20.01 (0.02) & 20.76 (0.03) & 20.10 (0.01) & 20.95 (0.02) & 19.38 (0.02) & 20.00 (0.03) \\
 V   & 15.20 (0.01) & 15.62 (0.03) & 19.56 (0.02) & 17.68 (0.01) & 19.08 (0.02) & 19.17 (0.01) & 19.47 (0.01) & 18.96 (0.01) & 19.70 (0.03) & 18.37 (0.02) & 19.66 (0.02) \\
 R   & 14.87 (0.01) & 15.26 (0.01) & 18.67 (0.01) & 16.81 (0.01) & 18.39 (0.01) & 18.64 (0.01) & 18.58 (0.02) & 18.19 (0.01) & 18.73 (0.02) & 17.74 (0.01) & 19.23 (0.01) \\
 I   & 14.54 (0.01) & 14.94 (0.01) & 17.76 (0.02) & 15.99 (0.02) & 17.80 (0.02) & 18.18 (0.02) & 17.59 (0.01) & 17.47 (0.01) & 17.68 (0.01) & 17.15 (0.02) & 18.96 (0.01) \\ [1ex] \hline
UVw2 & 18.34 (0.15) & 18.81 (0.13) &   -          &   -          &   -          &   -          &   -          &   -          &   -          &   -          & 20.31 (0.16) \\
UVm2 & 18.20 (0.18) & 18.61 (0.15) &   -          &   -          &   -          &   -          &   -          &   -          &   -          &   -          & 19.67 (0.16) \\
UVw1 & 16.92 (0.08) & 17.39 (0.07) &   -          &   -          &   -          &   -          &   -          &   -          &   -          &   -          & 19.45 (0.17) \\
 u   & 15.64 (0.05) & 16.01 (0.05) &   -          &   -          &   -          &   -          &   -          &   -          &   -          & 20.25 (0.20) & 19.05 (0.21) \\
 b   & 15.70 (0.07) & 16.19 (0.05) &   -          &   -          & 20.10 (0.27) & 19.84 (0.33) & 20.60 (0.34) &   -          & 20.70 (0.19) & 19.33 (0.20) & 20.17 (0.22) \\
 v   & 15.28 (0.07) & 15.65 (0.09) &   -          &   -          & 19.23 (0.22) & 19.09 (0.21) & 19.37 (0.21) &   -          & 19.73 (0.16) & 18.50 (0.22) & 19.98 (0.28) \\ [1ex] \hline
 J   & 14.12 (0.03) & 14.43 (0.03) &   -          & 15.05 (0.05) &   -          &   -          & 16.61 (0.15) & 16.75 (0.10) & 16.61 (0.15) & 16.58 (0.15) &  -           \\
 H   & 13.84 (0.04) & 14.02 (0.05) &   -          & 14.35 (0.05) &   -          &   -          & 15.85 (0.18) & 15.90 (0.19) & 15.83 (0.18) & 15.80 (0.17) &  -           \\ 
 Ks  & 13.71 (0.06) & 14.13 (0.07) &   -          & 14.18 (0.07) &   -          &   -          &   -          & 15.56 (0.24) & 15.73 (0.26) & 15.29 (0.17) &  -           \\ [1ex] \hline
\end{tabular}
\label{tab_seqstars}
\end{sidewaystable}

\end{document}